\documentclass[superscriptaddress,twocolumn,pre]{revtex4}

\usepackage{amsmath}
\usepackage{graphicx,color}

\newcommand{\be}{\begin{equation}}
\newcommand{\ee}{\end{equation}}
\newcommand{\bea}{\begin{eqnarray}}
\newcommand{\eea}{\end{eqnarray}}

\begin{document}
\sloppy


\title{Mass-radius relation of Newtonian self-gravitating Bose-Einstein condensates \\
with short-range interactions: II. Numerical results}

\author{Pierre-Henri Chavanis}
\affiliation{Laboratoire de Physique Th\'eorique (IRSAMC), CNRS and UPS, Universit\'e de Toulouse, France}
\author{Luca Delfini}
\affiliation{Laboratoire de Physique Th\'eorique (IRSAMC), CNRS and UPS, Universit\'e de Toulouse, France}

\begin{abstract}
We develop the suggestion that dark matter could be a Bose-Einstein condensate. We determine the mass-radius relation of a Newtonian self-gravitating Bose-Einstein condensate with short-range interactions described by the Gross-Pitaevskii-Poisson system. We numerically solve the equation of hydrostatic equilibrium describing the balance between the gravitational attraction and the pressure due to quantum effects (Heisenberg's uncertainty principle) and short-range interactions (scattering). We connect the non-interacting limit to the Thomas-Fermi limit. We also consider the case of attractive self-interaction. We compare the exact mass-radius relation obtained numerically with the approximate analytical relation obtained with a Gaussian ansatz. An overall good agreement is found.
\end{abstract}

\maketitle


\section{Introduction}

Several recent astrophysical observations of distant type Ia supernovae have revealed that the content of the universe is made of about $70\%$ of dark energy, $25\%$ of dark matter and $5\%$ of baryonic (visible) matter \cite{novae}.  Thus, the overwhelming preponderance of matter and energy in the universe is believed to be dark i.e. unobservable by telescopes.  The dark energy is responsible for the accelerated expansion of the universe. Its origin is mysterious and presumably related to the cosmological constant. Dark energy is usually interpreted as a vacuum energy and it behaves like a fluid with negative pressure. Dark matter also is mysterious. The suggestion that dark matter may constitute a large part of the universe was raised by Zwicky \cite{zwicky} in 1933. Using the virial theorem to infer the average mass of galaxies within the Coma cluster, he obtained a value much larger than the mass of luminous material. He realized therefore that some mass was ``missing'' in order to account for observations. This missing mass problem was confirmed later by accurate measurements of rotation curves of disc galaxies  \cite{rubin,persic}. The rotation curves of neutral hydrogen clouds in spiral galaxies measured from the Doppler effect are found to be roughly flat (instead of Keplerian) with a typical rotational velocity $v_\infty\sim 200\, {\rm km/s}$ up to the   maximum observed radius of about $50$ kpc. This mass profile is much more extended than the distribution of starlight which typically converges within $\sim 10$ kpc. This implies that galaxies are surrounded by an extended halo of dark matter whose mass $M(r)=r v_\infty^2/G$ increases linearly with radius \footnote{Some authors like Milgrom \cite{milgrom}  propose a modification of Newton's law (MOND theory) to explain the rotation curves of spiral galaxies without invoking dark matter.}. This can be conveniently modeled by an isothermal self-gravitating gas the density of which scales asymptotically as $r^{-2}$ \cite{chandra}.

The nature of dark matter (DM) is one of the most important puzzles in modern physics and cosmology. A wide ``zoology'' of exotic particles that could form dark matter has been proposed. In particular, many grand unified theories in particle physics predict the existence of various exotic  bosons (e.g. ultra-light bosons, axions, scalar neutrinos, neutralinos) that should be present in considerable abundance in the universe and comprise (part of) the cosmological missing mass \cite{primack,overduin}. Although the bosonic particles have never been detected in accelerator experiments, they are considered as leading candidates of dark matter and might play a significant role in the evolution and in the structure of the universe.

If dark matter is made of bosons, they probably have collapsed through
some sort of Jeans instability to form compact gravitating objects
such as boson stars. Boson stars were introduced by Kaup \cite{kaup}
and Ruffini \& Bonazzola \cite{rb} in the 1960s. Early works on
boson stars \cite{thirring,takasugi,breit,bij} were motivated by the
axion field that was proposed as a possible solution to the strong CP
problem in QCD. However, these works consider {mini boson stars}, like
axion black holes, with unrealistic small
masses. These mini boson stars could play a role,
however, if they exist in the universe in abundance or if the axion
mass is extraordinary small \footnote{String theory
generically predicts the existence of very light bosonic particles
down to masses of the order $10^{-33}{\rm eV}/c^2$.} leading to
macroscopic objects with a mass $M_{Kaup}$ comparable to the mass of
the sun (or even larger) \cite{mielke}. For example, axionic boson
stars could account for the mass of MACHOs (between $0.3$ and $0.8$
$M_{\odot}$) if the axions have a mass $m\sim 10^{-10}{\rm eV}/c^2$
\cite{mielkeschunck}.  Alternatively, the possibility that dark
matter could be made of {\it self-interacting} bosonic particles that
have formed boson stars with stellar masses was proposed by Colpi {\it
et al.} \cite{colpi} and their cosmological formation was discussed by
Bianchi {\it et al.}
\cite{bianchi} and Madsen \& Liddle
\cite{madsen}. The
presence of even a small repulsive self-interaction
$\frac{1}{4}\lambda |\phi|^4$ between bosons can considerably increase
the mass of the boson stars allowing therefore to consider larger
particle masses. For example, for $m\sim 1{\rm GeV}/c^2$ and
$\lambda\sim 1$, this mass is of the order of the solar mass
$M_{\odot}$, like in the case of white dwarf and neutron stars,
whereas, in
the absence of interaction ($\lambda=0$), $M_{Kaup}\sim 10^{-19}M_{\odot}$ for $m\sim 1{\rm GeV}/c^2$. Therefore, (mini) boson
stars could be the constituents of dark matter halos.

On the other hand, some authors
\cite{baldeschi,sin,leekoh,schunckpreprint,guzmanmatos,hu,peebles,goodman,arbey1,matosall,silverman,arbey,bohmer,fmt,sikivie,lee}
have proposed that dark matter halos themselves could be in the form
of gigantic self-gravitating Bose-Einstein condensates (with or
without self-interaction \footnote{If the bosons do not have
self-interaction, their mass must be extremely small, of the order of
$m\sim 10^{-24}\, {\rm eV}/c^2$, to reproduce the mass and size of
dark matter halos \cite{baldeschi,sin,hu,arbey1,silverman}. Ultralight
scalar fields like axions may have such small masses. Alternatively,
in the presence of self-interaction, the required mass of the bosons
can be increased up to $m\sim 1\, {\rm eV}/c^2$ to yield realistic
models of dark matter halos \cite{arbey,bohmer}.}) described by a
single wave function $\psi({\bf r},t)$. In the Newtonian limit, which
is relevant at the galactic scale, the evolution of this wave function
is governed by the Gross-Pitaevskii-Poisson (GPP) system
\cite{bohmer}. Using the Madelung \cite{madelung} transformation, the
GP equation turns out to be equivalent to hydrodynamic (Euler)
equations involving a barotropic isotropic pressure due to short-range
interactions (scattering) and an anisotropic quantum pressure arising
from the Heisenberg uncertainty principle. At large scales, quantum
effects are negligible and one recovers the classical hydrodynamic
equations of cold dark matter (CDM) models which are remarkably
successful in explaining the large-scale structure of the universe
\cite{ratra}. At small-scales, gravitational collapse is prevented by
the repulsive scattering or by the uncertainty principle. This may be
a way to solve the problems of the CDM model such as the cusp problem
\cite{observations} and the missing satellite problem
\cite{satellites}.

The approach developed in our series of papers
\cite{paper1,paper3,cosmobec} is essentially theoretical and aims at a
general study of the basic equations of the problem. In previous
works, two important limits have been considered.  Ruffini \&
Bonazzola \cite{rb} studied boson stars without self-interaction
described by the Schr\"odinger-Poisson system (in the Newtonian
limit). The equilibrium state results from a balance between
gravitational attraction and quantum pressure arising from the
uncertainty principle. On the other hand, B\"ohmer \& Harko
\cite{bohmer} studied self-gravitating BECs with short-range
interactions described by the Gross-Pitaevskii-Poisson system and
considered the Thomas-Fermi (TF) limit in which quantum pressure is
negligible. In that case, the equilibrium state results from a balance
between gravitational attraction and repulsive scattering (for
positive scattering lengths $a>0$). The TF approximation is valid for
$GN^2m^3a/\hbar^2\gg 1$ \cite{paper1} where $N$ is the number of
particles. Since there is some
indetermination on the values of the mass $m$ and scattering length
$a$ of the bosons, it may be conceptually interesting to treat the
general case in detail and connect these two asymptotic limits. This
program was started in Paper I \cite{paper1} using an analytical
approach based on a Gaussian ansatz and a mechanical analogy. For the
sake of completeness, we also considered the case of attractive
short-range interactions (i.e. negative scattering lengths $a<0$) and
reported the existence of a maximum mass $M_{max}$ above which no
equilibrium state exists. We argued that attractive self-interaction
could accelerate the formation of structures in the universe. In this
paper, we shall test the validity of these analytical results by
determining the exact steady state of the GPP system numerically.

The paper is organized as follows. In Sec. \ref{sec_formal}, we
recapitulate the basic equations of the problem and specifically
mention the non-interacting case and the Thomas-Fermi limit. In
Sec. \ref{sec_numerical}, we explain our numerical procedure to solve
the quantum equation of hydrostatic equilibrium which is equivalent to
the steady state of the GPP system. In Sec. \ref{sec_fa}, we plot the
fundamental curves (radius, energy, density,...) as a function of the
mass in the case where the value of the scattering length is
prescribed. Finally, in Sec. \ref{sec_funmassfixed}, we plot these
curves as a function of the scattering length for a fixed value of the
total mass. We also compare the exact relations obtained numerically
with the approximate analytical relations obtained in Paper I and find
an overall good agreement. In Appendix \ref{sec_na},
we make numerical applications in order to compare our results
with real astrophysical objects and determine the validity of our
assumptions.

\section{The Gross-Pitaevskii-Poisson system}
\label{sec_formal}

\subsection{The Madelung transformation}
\label{sec_madelung}

A Newtonian self-gravitating BEC is described by the  Gross-Pitaevskii-Poisson system \cite{bohmer,paper1}:
\begin{equation}
\label{m1}
i\hbar \frac{\partial\psi}{\partial t}=-\frac{\hbar^2}{2m}\Delta\psi+m(\Phi+gNm |\psi|^2)\psi,
\end{equation}
\begin{equation}
\label{m2}
\Delta\Phi=4\pi G Nm |\psi|^2,
\end{equation}
with $g={4\pi a\hbar^2}/{m^3}$ where $a$ is the s-scattering length (we allow $a$ to be positive or negative) \cite{revuebec}. We write the wave function in the form $\psi({\bf r},t)=A({\bf r},t)e^{iS({\bf r},t)/\hbar}$ where $A$ and $S$ are real, and  make the Madelung \cite{madelung} transformation
\begin{equation}
\label{m3}
\rho=Nm|\psi|^2=NmA^2, \qquad  {\bf u}=\frac{1}{m}\nabla S,
\end{equation}
where $\rho({\bf r},t)$ is the density field and ${\bf u}({\bf r},t)$ the velocity field. The total mass of the configuration is $M=Nm=\int\rho\, d{\bf r}$. We note that the flow is irrotational since $\nabla\times {\bf u}={\bf 0}$.  With this transformation, it can be shown that the Gross-Pitaevskii equation (\ref{m1}) is equivalent to the barotropic Euler equations with an additional term called the quantum potential (or quantum pressure). Indeed, one obtains the set of equations
\begin{equation}
\label{m4}
\frac{\partial\rho}{\partial t}+\nabla\cdot (\rho {\bf u})=0,
\end{equation}
\begin{equation}
\label{m5}
\frac{\partial {\bf u}}{\partial t}+({\bf u}\cdot \nabla){\bf u}=-\frac{1}{\rho}\nabla p-\nabla\Phi-\frac{1}{m}\nabla Q,
\end{equation}
with
\begin{equation}
\label{m6}
Q=-\frac{\hbar^2}{2m}\frac{\Delta \sqrt{\rho}}{\sqrt{\rho}},
\end{equation}
and
\begin{equation}
\label{m7}
p=\frac{2\pi a\hbar^2}{m^3}\rho^{2}.
\end{equation}
The pressure relation (\ref{m7}) corresponds to a polytropic equation of state of the form
\begin{equation}
\label{m8}
p=K\rho^{\gamma},\qquad \gamma=1+\frac{1}{n},
\end{equation}
with polytropic index $n=1$ (i.e. $\gamma=2$) and polytropic constant $K={2\pi a\hbar^2}/{m^3}$.

{\it Remark:} If dark matter is treated as a collisionless fluid (as in standard
models \cite{peeblesbook,ratra}), the evolution of this fluid is fundamentally
described by the Vlasov-Poisson system. The hydrodynamic equations
without pressure ($\hbar=p=0$) can be obtained if we assume the
existence of a single fluid velocity at every spatial position and
they cease to be valid after the first time of crossing where
multi-streaming generates a range of particles velocities through a
given point \cite{vergassola}. In that case, the evolution of the
Vlasov-Poisson system becomes very complex and exhibits phenomena of
nonlinear Landau damping and phase mixing \cite{kandrup}. The
derivation of fluid equations in that context becomes difficult (it 
amounts to closing the infinite hierarchy of Jeans equations) even if we
may argue that a form of local thermodynamical equilibrium and an effective
pressure arise as a result of violent relaxation
\cite{csr}. Alternatively, if dark matter is a self-gravitating BEC,
the hydrodynamic equations (\ref{m4})-(\ref{m5}) are exact for all
times since they are rigorously equivalent to the GP equation
(\ref{m1}). In that case, the pressure $p$ is due to short-range
interactions and non-ideal effects, not to thermal effects
\cite{paper1}. There is also a quantum pressure. These pressure terms prevent multi-streaming and the formation of
singularities (caustics), and regularize the dynamics at small scales
\cite{kpz}. Therefore, BEC dark matter behaves like a fluid, contrary
to standard CDM that is essentially collisionless.

\subsection{The time independent GP equation}
\label{sec_steady}

If we consider a wave function of the form
\begin{equation}
\label{s1}
\psi({\bf r},t)=A({\bf r})e^{-i\frac{Et}{\hbar}},
\end{equation}
we obtain the  time-independent GP equation
\begin{equation}
\label{s2}
-\frac{\hbar^2}{2m}\Delta\psi+m(\Phi+g N m\psi^2)\psi=E\psi,
\end{equation}
where $\psi({\bf r})\equiv A({\bf r})$ is real and $\rho({\bf r})=Nm\psi^2({\bf r})$. The foregoing equation can be rewritten
\begin{equation}
\label{s3}
m\Phi+m g\rho-\frac{\hbar^2}{2m}\frac{\Delta\sqrt{\rho}}{\sqrt{\rho}}=E,
\end{equation}
or, equivalently,
\begin{equation}
\label{s4}
m\Phi+m g \rho+Q=E.
\end{equation}
In paper I, we have shown that this
equation can be obtained from a variational principle, by minimizing
the energy functional $E_{tot}$ at fixed mass $M$. Combined with the
Poisson equation (\ref{m2}), we obtain an eigenvalue equation for the
wave function $\psi({\bf r})$ where the eigenvalue $E$ is the energy.
In the following, we shall be interested in the fundamental eigenmode
corresponding to the smallest value of $E$. For this mode, the wave
function $\psi(r)$ is spherically symmetric and has no node so that
the density profile decreases monotonically with the distance.

\subsection{Hydrostatic equilibrium}
\label{sec_hydro}

The time-independent solution (\ref{s3}) can also be obtained from the Euler equations (\ref{m4})-(\ref{m6}) since they are equivalent to the GP equation. The steady state of the quantum Euler equation (\ref{m5}) obtained by taking $\partial_t=0$ and ${\bf u}={\bf 0}$ satisfies
\begin{equation}
\label{h1}
\nabla p+\rho\nabla\Phi-\frac{\hbar^2\rho}{2m^2}\nabla \left (\frac{\Delta\sqrt{\rho}}{\sqrt{\rho}}\right )={\bf 0}.
\end{equation}
This is similar to the condition of hydrostatic equilibrium with an additional quantum potential. It describes the balance between pressure due to short-range interactions (scattering), quantum pressure (Heisenberg principle) and gravity. This equation is equivalent to Eq. (\ref{s3}). Indeed, integrating Eq. (\ref{h1}) using Eq. (\ref{m7}), we obtain Eq. (\ref{s3}) where the eigenenergy  $E$ appears as a constant of integration. Combining Eq. (\ref{h1}) with the Poisson equation (\ref{m2}), we obtain the fundamental equation of hydrostatic equilibrium with quantum effects
\begin{equation}
\label{h2}
-\nabla\cdot \left (\frac{\nabla p}{\rho}\right )+\frac{\hbar^2}{2m^2}\Delta \left (\frac{\Delta\sqrt{\rho}}{\sqrt{\rho}}\right )=4\pi G\rho.
\end{equation}
For the polytropic equation of state (\ref{m7}), we get
\begin{equation}
\label{h3}
-\frac{4\pi a \hbar^2}{m^3}\Delta\rho+\frac{\hbar^2}{2m^2}\Delta \left (\frac{\Delta\sqrt{\rho}}{\sqrt{\rho}}\right )=4\pi G\rho.
\end{equation}
There are three important limits to consider. 

The {\it non-interacting case} corresponds to $g=a=0$. This is the
situation first studied by Ruffini \& Bonazzola \cite{rb} and
revisited by Membrado {\it et al.} \cite{membrado} with another
method. In that case, Eq. (\ref{h3}) reduces to
\begin{equation}
\label{h4}
\frac{\hbar^2}{2m^2}\Delta \left (\frac{\Delta\sqrt{\rho}}{\sqrt{\rho}}\right )=4\pi G\rho.
\end{equation}
It describes the balance between attractive gravity and repulsive quantum pressure (Heisenberg principle). This equation must be solved numerically. It is found \cite{rb,membrado} that the density profile decays smoothly to infinity and that the radius containing $99\%$ of the mass is given by $R_{99}=9.9\hbar^2/GMm^2$. 

For $a>0$, we can make the {\it Thomas-Fermi approximation} which amounts to neglecting the quantum potential. This is the limit considered by B\"ohmer \& Harko \cite{bohmer}. In that case, Eq. (\ref{h3}) becomes
\begin{equation}
\label{h5}
\Delta\rho+\frac{Gm^3}{a\hbar^2}\rho=0.
\end{equation}
It describes the balance between attractive gravity and repulsive short-range interactions (scattering). This equation is equivalent to the Lane-Emden equation for a polytrope of index $n=1$ \cite{chandra}. It has the analytical solution $\rho(r)=(\rho_0 R/\pi r)\sin\left (\pi r/{R}\right )$ where $R=\pi (a\hbar^2/Gm^3)^{1/2}$ is the radius of the configuration (independent on the mass $M$) and $\rho_0=\pi M/4R^3$ is the central density. 

For $a<0$, the TF approximation leads to collapse since it only keeps the effects of attractive gravity and attractive short-range interactions (scattering). If we want to obtain equilibrium states (for $M<M_{max}$ \cite{paper1}), we have to solve the complete Eq. (\ref{h3}) expressing the balance between repulsive quantum pressure (Heisenberg principle) and attractive short-range interactions (scattering) and gravity. We could also consider the {\it non-gravitational limit}. In that case, 
Eq. (\ref{h3}) reduces to
\begin{equation}
\label{h5nwe}
\frac{4\pi |a| \hbar^2}{m^3}\Delta\rho+\frac{\hbar^2}{2m^2}\Delta \left (\frac{\Delta\sqrt{\rho}}{\sqrt{\rho}}\right )=0.
\end{equation}
It describes the balance between repulsive quantum pressure (Heisenberg principle) and  attractive short-range interactions (scattering). However, in Paper I, we have shown that such equilibria are unstable.

\section{Numerical solution}
\label{sec_numerical}

\subsection{The fundamental differential equation}
\label{sec_fundamental}

The structure of a self-gravitating BEC with short-range interactions is determined by the condition of hydrostatic equilibrium (\ref{h3}). In the general case, this equation has to be solved numerically. To that purpose, it is convenient to introduce dimensionless variables \cite{membrado}. We introduce the lengthscale
\begin{equation}
\label{f1}
b=\frac{\hbar^2}{2GMm^2},
\end{equation}
which  basically corresponds to the radius of a self-gravitating BEC in the absence of short-range interaction \cite{rb}.  We then define the  dimensionless position ${\bf x}$ and the dimensionless density profile $n({\bf x})$ by
\begin{equation}
\label{f2}
{\bf x}=\frac{{\bf r}}{b},\qquad n({\bf x})=\frac{4\pi b^3}{Nm}\rho({\bf r}).
\end{equation}
We also introduce the dimensionless eigenenergy $\epsilon$ and the dimensionless gravitational potential $\phi({\bf x})$ through the relations
\begin{eqnarray}
\label{f3}
\epsilon=\frac{2mb^2}{\hbar^2}E,\qquad \phi({\bf x})=\frac{2m^2b^2}{\hbar^2}\Phi({\bf r}).
\end{eqnarray}
Finally, we introduce the dimensionless parameter
\begin{equation}
\label{f4}
\chi=\frac{2Na}{b}=\frac{4GM^2ma}{\hbar^2}.
\end{equation}
It is equal to the ratio between $N$ times the scattering length $a$ and the typical radius $b$ of a self-gravitating BEC without interaction. The non-interacting limit corresponds to $\chi\ll 1$ and the TF limit corresponds to $\chi\gg 1$ \cite{paper1}.

In terms of these dimensionless variables, the equation of hydrostatic equilibrium (\ref{h3}) can be rewritten
\begin{equation}
\label{f5}
\Delta \left (\frac{\Delta\sqrt{n}}{\sqrt{n}}\right )-\chi\Delta n=n.
\end{equation}
The density must satisfy the normalization condition
\begin{eqnarray}
\label{f6}
\int n({\bf x})\, d{\bf x}=4\pi.
\end{eqnarray}
Finally, the normalized eigenenergy $\epsilon$ can be obtained from the steady state equation (\ref{s3}) leading to
\begin{eqnarray}
\label{f7}
\phi+\chi n-\frac{\Delta\sqrt{n}}{\sqrt{n}}=\epsilon,
\end{eqnarray}
where $n({\bf x})$ is the solution of Eq. (\ref{f5}) and $\phi({\bf x})$ is given by
\begin{eqnarray}
\label{f8}
\phi({\bf x})=-\frac{1}{4\pi}\int\frac{n({\bf x}')}{|{\bf x}-{\bf x}'|}\, d{\bf x}',
\end{eqnarray}
equivalent to the Poisson equation (\ref{m2}). For a spherically symmetric distribution, Eq. (\ref{f5}) reduces to the ordinary differential equation
\begin{eqnarray}
\label{f9}
n''''+\frac{4}{x}n'''-\frac{10 n'n''}{n x}+\frac{6n'^3}{n^2x}-\frac{3n'''n'}{n}-\frac{2n''^2}{n}\nonumber\\
+\frac{7n'^2n''}{n^2}-\frac{3n'^4}{n^3}-2n\chi\left (n''+\frac{2n'}{x}\right )=2n^2,
\end{eqnarray}
where the prime $'$ denotes $d/dx$. The normalization condition (\ref{f6}) takes the form
\begin{eqnarray}
\label{f10}
\int_{0}^{+\infty} n x^2\, dx=1,
\end{eqnarray}
and the steady state equation (\ref{f7}) becomes
\begin{eqnarray}
\label{f11}
\epsilon=\phi+\chi n-\frac{n''}{2n}-\frac{n'}{nx}+\frac{n'^2}{4n^2}.
\end{eqnarray}
The density behaves near the origin like $n(x)\simeq n_0+n_2x^2+...$. Taking the limit $x\rightarrow 0$ in Eq. (\ref{f11}), and using the field equation (\ref{f8}), we find that
\begin{eqnarray}
\label{f12}
\epsilon=-\int_{0}^{+\infty}n(x)x\, dx+\chi n_0-\frac{3n_2}{n_0}.
\end{eqnarray}
The procedure to determine the equilibrium state is now clear. We first have to solve the differential equation (\ref{f9}) with the boundary conditions
$n(0)=n_0$, $n'(0)=n'''(0)=0$ and $n''(0)=2n_2$. The constants $n_0$ and
$n_2$ have to be determined so as to yield  a physical density profile at infinity and satisfy the normalization condition (\ref{f10}). Finally, the eigenenergy $\epsilon$ is given by Eq. (\ref{f12}).

To solve this problem, it is convenient to introduce the function $f(x)=n(x)/n_0$
and make the change of variables $X=n_0^{1/4}x$ and $\mu=n_0^{1/2}\chi$. We thus have to
solve the ordinary differential equation
\begin{eqnarray}
\label{f13}
f''''+\frac{4}{X}f'''-\frac{10 f'f''}{f X}+\frac{6f'^3}{f^2X}-\frac{3f'''f'}{f}-\frac{2f''^2}{f}\nonumber\\
+\frac{7f'^2f''}{f^2}-\frac{3f'^4}{f^3}-2f\mu\left (f''+\frac{2f'}{X}\right )=2f^2,
\end{eqnarray}
with the boundary conditions
\begin{eqnarray}
\label{f14}
f(0)=1, \quad f'(0)=f'''(0)=0,\quad f''(0)=\frac{2n_2}{n_0^{3/2}}\equiv A_2.\nonumber\\
\end{eqnarray}
For given $\mu$, the constant $A_2$ is determined so as to yield a physical density profile at infinity (see below). Then, the constant $n_0$ is determined by the normalization condition (\ref{f10}) which becomes
\begin{eqnarray}
\label{f15}
\frac{1}{n_0^{1/4}}=\int_{0}^{+\infty} f(X) X^2\, dX.
\end{eqnarray}
Once $A_2$ and $n_0$ are known, we can obtain the coefficient $n_2=A_2 n_0^{3/2}/2$, the parameter $\chi=\mu/ n_0^{1/2}$ and the normalized eigenenergy
\begin{eqnarray}
\label{f16}
\epsilon=-\sqrt{n_0}\int_{0}^{+\infty}f(X)X\, dX+\sqrt{n_0}\mu-\frac{3n_2}{n_0}.
\end{eqnarray}

\subsection{The numerical procedure}
\label{sec_procedure}

\begin{figure}[!h]
\begin{center}
\includegraphics[clip,scale=0.3]{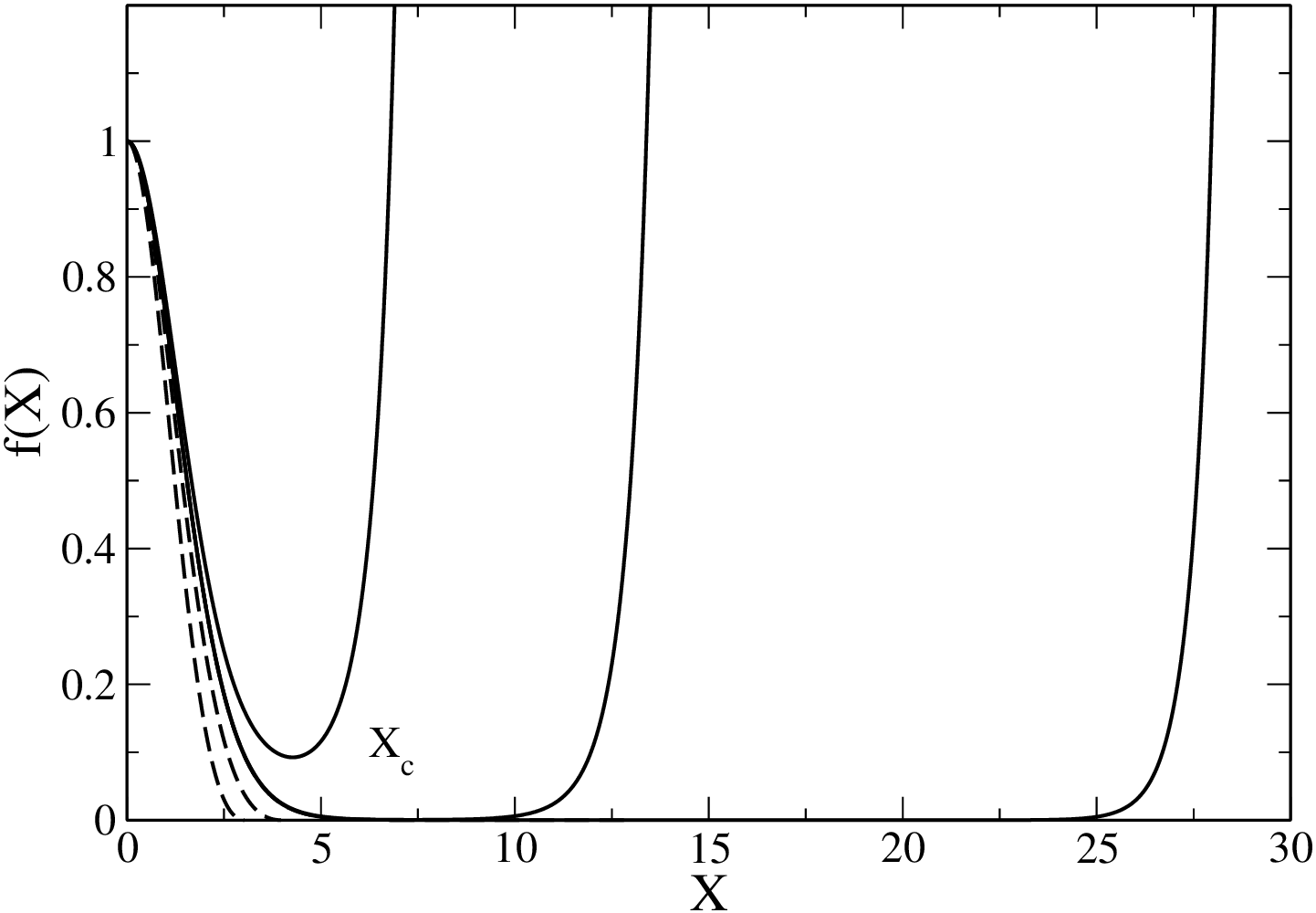}
\caption{Illustration of the shooting problem for determining the optimal value of $(A_2)_*$. This example  corresponds to $\chi=\mu=0$ but it is representative of more general cases.}
\label{explication}
\end{center}
\end{figure}

\begin{figure}[!h]
\begin{center}
\includegraphics[clip,scale=0.3]{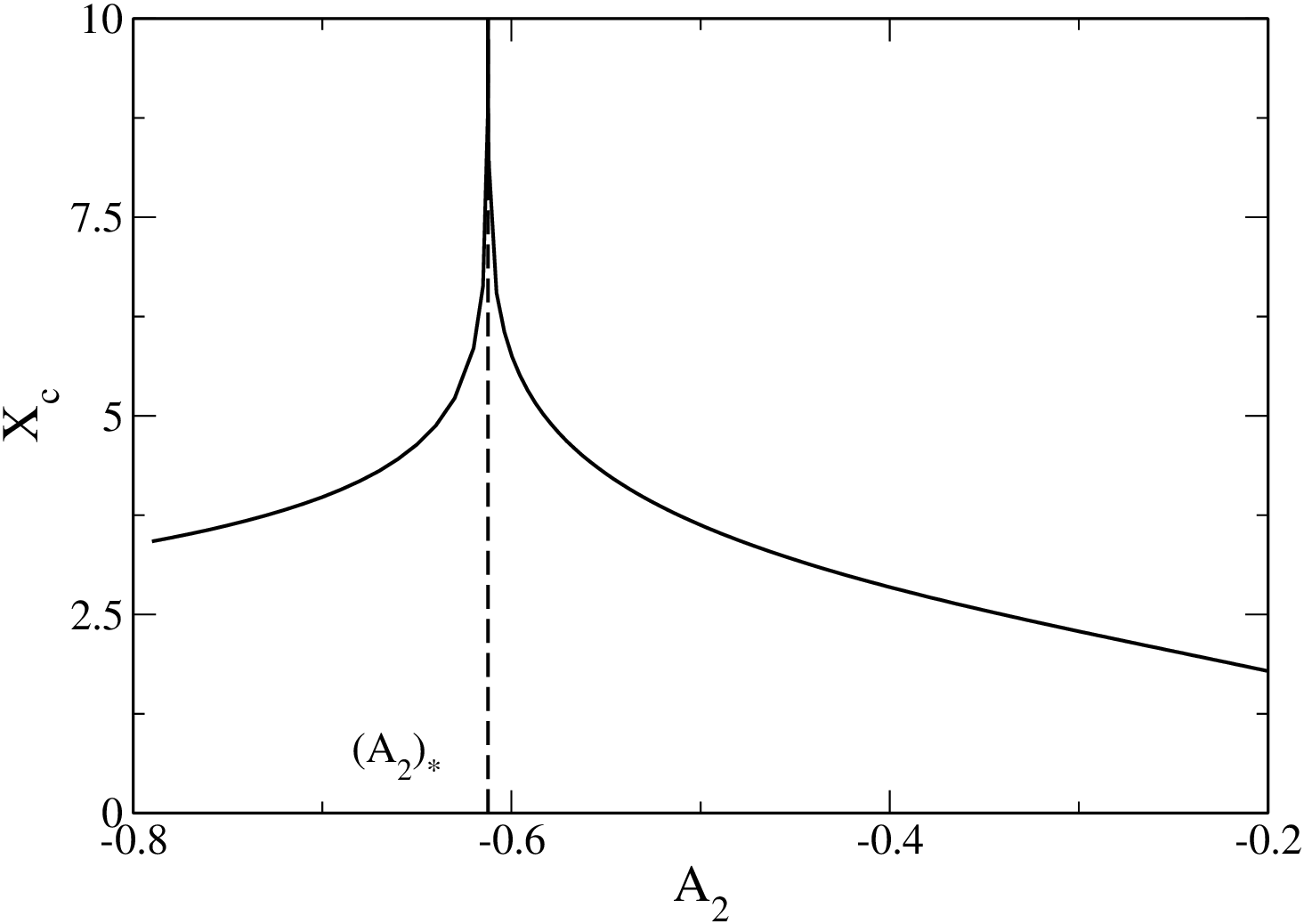}
\caption{Evolution of $X_c$ as $A_2$ approaches $(A_2)_*$. We find that the convergence is logarithmic $X_{c}\propto \ln |A_2-(A_2)_*|$ so that a huge precision on the value of $A_2$ is needed to obtain a large value of $X_{max}$.  This example corresponds to $\chi=\mu=0$ but it is representative of more general cases.}
\label{Xmax}
\end{center}
\end{figure}

The numerical procedure that was followed to determine the proper value of $A_2$, denoted   $(A_2)_*$, can be understood from the example shown in Fig. \ref{explication}. If we solve the differential equation (\ref{f13}) with $A_2>(A_2)_*$ (full lines), the density profile reaches a minimum $f_{min}$  at $X=X_{c}$ before increasing indefinitely. As $A_2$ approaches $(A_2)_*$ from above, the point $X_c$ is pushed further and further away while $f_{min}$ decreases. Therefore, the divergence of the density occurs at larger and larger radii. On the other hand, if we solve the differential equation with $A_2<(A_2)_*$ (dashed lines), the density profile decreases until a point $X=X_{c}$ at which the program breaks down because the density achieves too small values ($<10^{-11}$). As $A_2$ approaches $(A_2)_*$ from below, the point $X_c$ is pushed further and further away so that the ``break down'' occurs at larger and larger radii. Ideally, if we could start exactly from $(A_2)_*$, the point $X_c$ would be rejected to $+\infty$ and the density profile would gently decrease towards zero at infinity. In practice, it is impossible to obtain the ``exact'' value of $(A_2)_*$. Furthermore, it is shown in Fig. \ref{Xmax} that the value of $X_c$ increases very slowly (logarithmically) with the distance  $|A_2-(A_2)_*|$ to the exact value, so that a huge precision on the value of $A_2$ is needed to obtain large values of $X_{c}$.  Let us call $X_{max}$ the largest value of $X_c$ that we have been able to obtain numerically. For $\chi=\mu=0$, we have determined $(A_2)_{*}$ with a precision of the order of $10^{-12}$, i.e. $(A_2)_*\simeq -0.612386937160$, to get $X_{max}\simeq 18.66$. The normalized central density and the normalized eigenenergy are found to be $n_0\simeq 6.911\, 10^{-3}$ and $\epsilon\simeq -8.138\, 10^{-2}$. Finally, the radius $x$ containing $99\%$ of the mass is $\Lambda_{99}\simeq 19.89$. These values are consistent with those obtained in \cite{membrado}.

\begin{figure}[!h]
\begin{center}
\includegraphics[clip,scale=0.3]{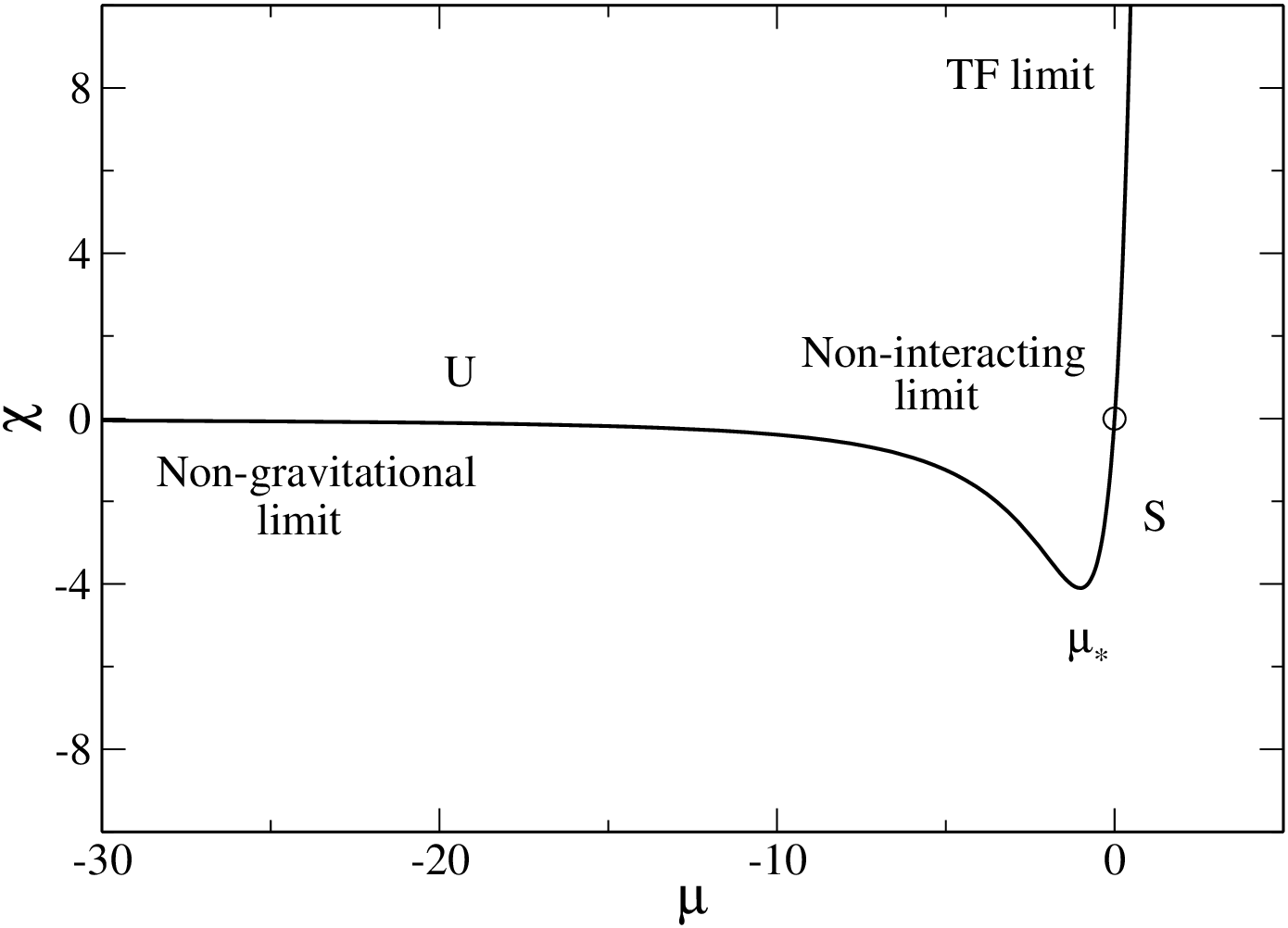}
\caption{Evolution of the parameter $\chi$ as a function of $\mu$.}
\label{chi}
\end{center}
\end{figure}

More generally, we have solved this shooting problem for several values of $\mu$. The series of equilibria, that will be studied in the following sections, is parameterized by $\mu$ taking values between $-\infty$ and $+\infty$. Figure \ref{chi} shows that the parameter $\chi$ is not a monotonic function of $\mu$ in the region corresponding to negative scattering lengths ($a<0$). Indeed, $\chi$ starts from $0^-$ when $\mu\rightarrow -\infty$, reaches a minimum value $\chi_*\simeq -4.100$ at $\mu_*\simeq -1.000$, returns to $0$ at $\mu=0$  and finally increases to $+\infty$ when $\mu\rightarrow +\infty$. The TF limit corresponds to $\mu\rightarrow +\infty$, the non-interacting limit corresponds to $\mu=0$ and the non-gravitational limit corresponds to $\mu\rightarrow -\infty$. The non-monotonicity of $\chi(\mu)$ is associated with an instability. This is related to the Poincar\'e theorem  (see, e.g., \cite{katz,ijmpb}) since the parameter $\chi$ plays the role of the mass for a given scattering length $a$. Since the stable equilibrium configurations are minima of energy at fixed mass \cite{paper1}, a direct application of the Poincar\'e theorem (see Sec. \ref{sec_poincare}) implies that a change of stability occurs at a turning point of mass, hence of $\chi$. Since we know that the system is stable in the TF limit (because it is equivalent to a polytrope of index $\gamma=2$ larger than the critical index $\gamma_c=4/3$ \cite{bt}), we conclude from the Poincar\'e theorem that all the configurations with $\mu>\mu_*$ are dynamically stable (S) while all the configurations with $\mu<\mu_*$ are unstable (U). In particular, the system is unstable in the non-gravitational limit. The same conclusions have been reached in Paper I based on the analytical model.

\subsection{The dimensionless parameters}
\label{sec_dim}

As we have seen, the structure of the problem depends on a single
control parameter $\chi$ given by Eq. (\ref{f4}). There are, however,
two ways to present the results. In the first case, we assume that the
scattering length $a$ is fixed and we study how the physical
parameters like the radius, the energy, the density,... depend on the
mass.  This is certainly the most relevant representation for
astrophysical problems. In particular, we shall determine the
mass-radius relation $M(R)$ of boson stars as was done in the past for
white dwarfs \cite{chandra} and neutron stars \cite{shapiro}. In the
second case, we assume that the mass $M$ is fixed and we study how the
physical parameters depend on the scattering length. In particular, we
shall determine the radius versus scattering length relation
$R(a)$. This representation can also be of interest. Therefore, we
shall treat the two situations successively. In each case, the
physical parameters (mass, radius, scattering length,...) must be
properly normalized as explained below.

$\bullet$ Let us first consider the case where the mass $M$ is fixed. Using the definitions of Paper I, we normalize the scattering length by $a_Q=\hbar^2/GM^2m$, the radius by $R_Q=\hbar^2/GMm^2$, the density by $\rho_Q=M/R_Q^3=G^3M^4m^6/\hbar^6$, the total energy by $E_Q=GM^2/R_Q=G^2M^3m^2/\hbar^2$ and the eigenenergy by $E_Q'=GMm/R_Q=G^2M^2m^3/\hbar^2$. In terms of these variables, we obtain $a/a_Q=\chi/4$, $r/R_Q=x/2$, $R_{99}/R_Q=\Lambda_{99}/2$, $E/E_Q'=2\epsilon$ and $\rho(r)/\rho_Q=(2/\pi)n(x)$. The TF limit is valid for $a\gg a_Q$ and the non-interacting limit for $a\ll a_Q$.

$\bullet$ Let us now consider the case where the scattering length $a$ is fixed. We normalize the mass by $M_a=\hbar/\sqrt{Gm|a|}$, the radius by $R_a=(|a|\hbar^2/Gm^3)^{1/2}$, the density by $\rho_a=M_a/R_a^3=Gm^4/a^2\hbar^2$, the total energy by $E_a=GM_a^2/R_a=\hbar(Gm)^{1/2}/|a|^{3/2}$ and the eigenenergy by $E_a'=GM_a m/R_a=Gm^2/|a|$. In terms of these variables, we have $M/M_a=\sqrt{|\chi|}/2$, $r/R_a=x/\sqrt{|\chi|}$, $R_{99}/R_a=\Lambda_{99}/\sqrt{|\chi|}$, $E/E_a'=\epsilon |\chi|/2$ and $\rho(r)/\rho_a=\chi^2 n(x)/8\pi$. The TF limit is valid for $M\gg M_a$ and the non-interacting limit for $M\ll M_a$.

We can now plot the fundamental curves in the two situations described
above. The procedure is the following. For a given value of $\mu$, we
determine successively $(A_2)_*$, $n_0$, $n_2$, $\chi$ and
$\epsilon$. Then, we obtain the density profile $n(x)$. From the
density profile $n(x)$, we obtain the value $\Lambda_{99}$ of the
radius $x$ containing $99\%$ of the mass. Finally, by varying $\mu$
between $-\infty$ and $+\infty$, we obtain the fundamental curves
(parametrized by $\mu$) for a fixed value of the scattering length or
for a fixed value of the total mass by using the parameters defined
above. The quantities that appear in these curves
(mass, radius, density, energy,...) are dimensionless so that we are
not required to specify the characteristics of the bosons (that are
necessarily uncertain). In this sense, our approach is very general and
can model different systems such as (mini) boson stars and galactic
halos. In Appendix \ref{sec_na}, we make numerical applications
in order to compare our results with real astrophysical
objects.

\section{The fundamental curves for a fixed value of the scattering length}
\label{sec_fa}

\subsection{The mass-radius relation}
\label{sec_mraa}

\begin{figure}[!h]
\begin{center}
\includegraphics[clip,scale=0.3]{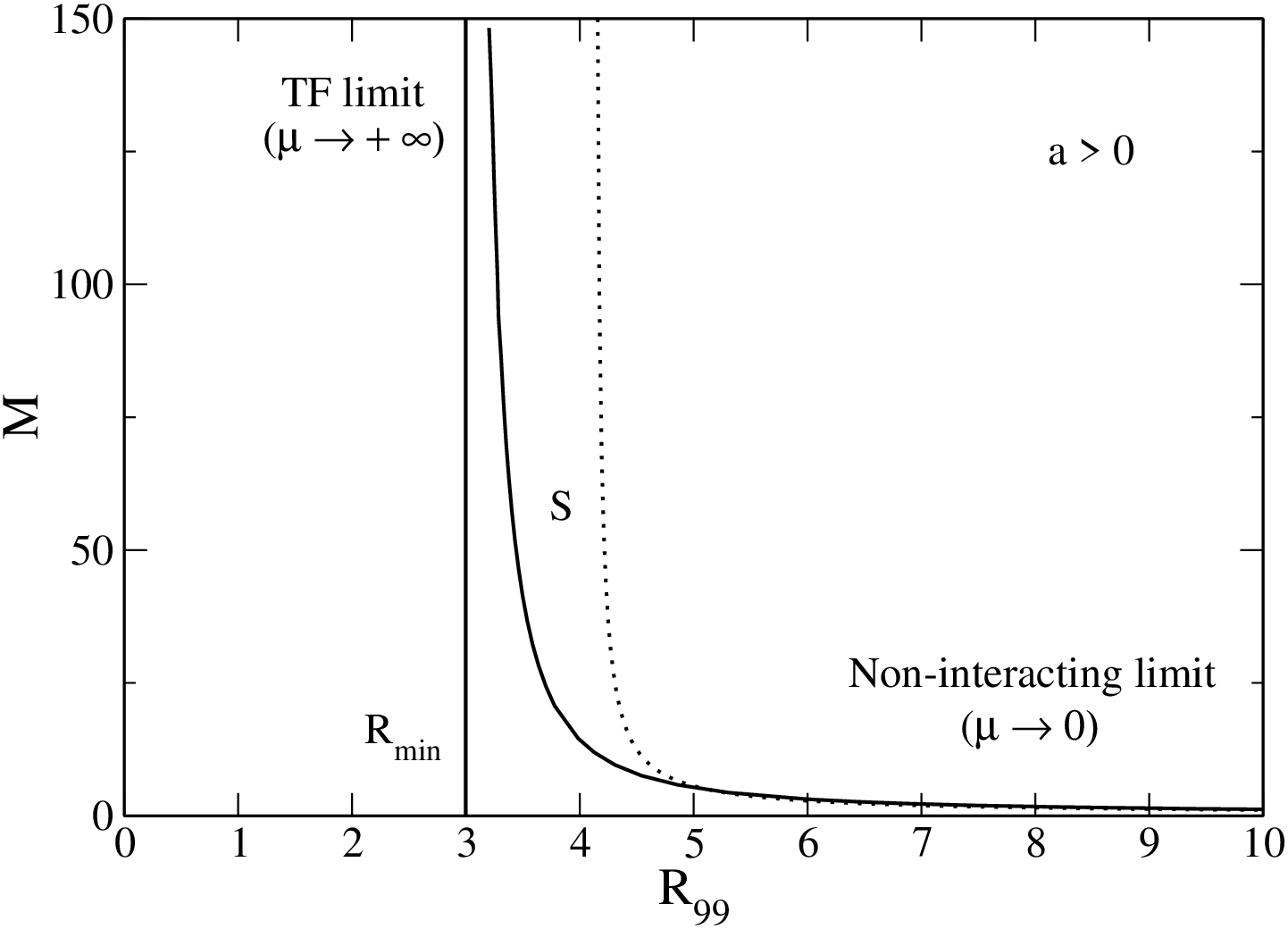}
\caption{${M}$ as a function of ${R}_{99}$ for fixed $a>0$. The mass is normalized by $M_a$ and the radius by $R_a$. In the non-interacting limit $M\rightarrow 0$, we get $M\sim 9.946/R_{99}$. In the TF limit $M\rightarrow +\infty$, we obtain $R_{99}\rightarrow 2.998$.  The system is always stable.  The dotted line corresponds to the approximate analytical mass-radius relation $M=2\sigma R/(\nu R^2-6\pi\zeta)$ (with $R_{99}=2.38167R$)  based on the Gaussian ansatz \cite{paper1}. The radius is given as a function of the mass by $R=(\sigma/\nu M)(1+\sqrt{1+6\pi\zeta\nu M^2/\sigma^2})$. In the non-interacting limit $M\rightarrow 0$, we get $M\sim 2\sigma/\nu R$ i.e. $M^{Gauss}\sim 8.955/R_{99}$ and in the TF limit $M\rightarrow +\infty$, we get $R\rightarrow (6\pi\zeta/\nu)^{1/2}$ i.e. $R_{99}^{Gauss}\rightarrow 4.125$. The analytical mass-radius relation  has the same qualitative shape as the numerical curve and provides a good quantitative agreement in the non-interacting limit. The agreement is less good in the TF limit where the density profile sensibly differs from a Gaussian.}
\label{M-R-chi-pos-part1}
\end{center}
\end{figure}

\begin{figure}[!h]
\begin{center}
\includegraphics[clip,scale=0.3]{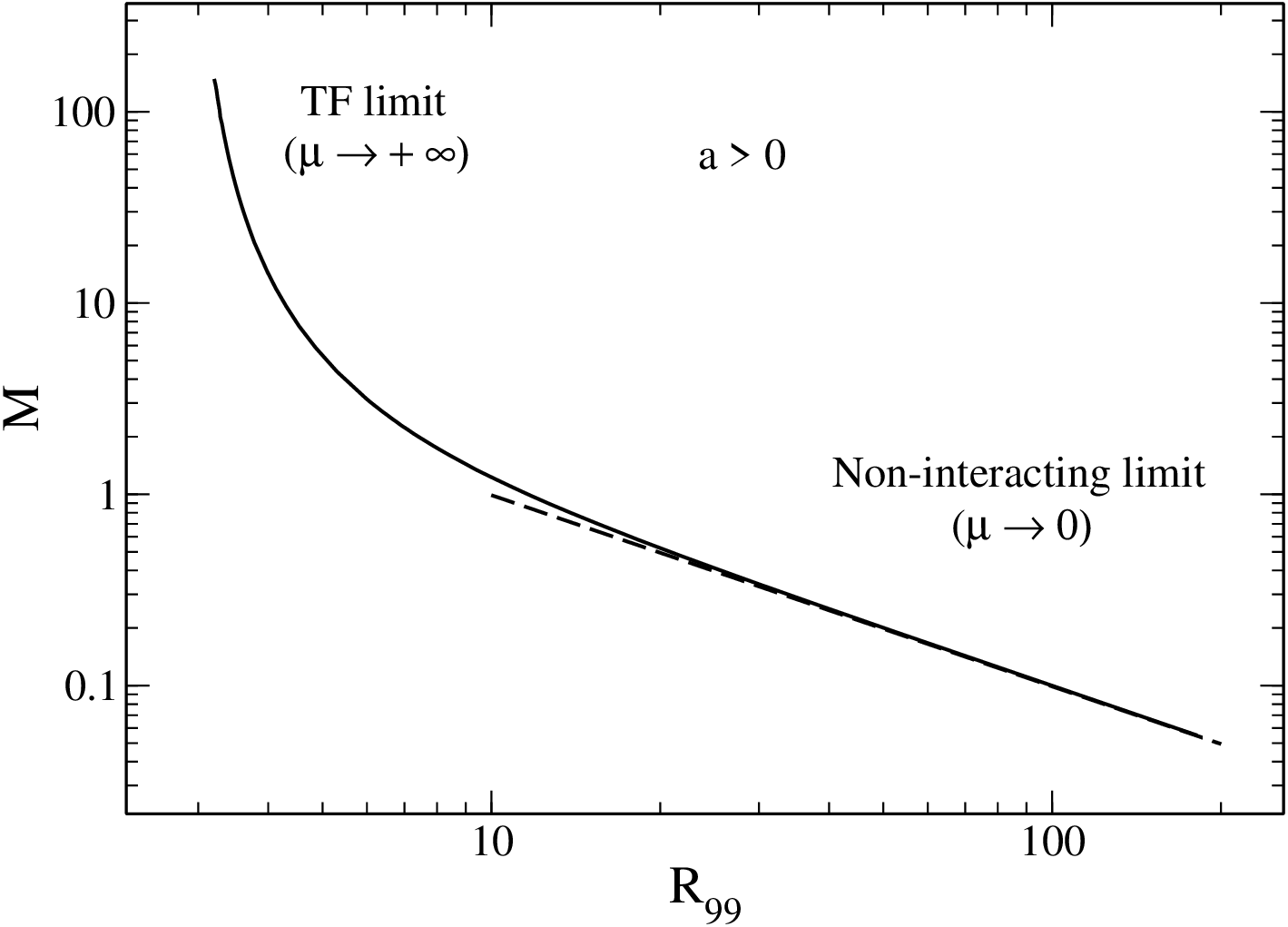}
\caption{${M}$ as a function of ${R}_{99}$ in log-log plot for fixed $a>0$. The dashed line corresponds to ${M}\sim 9.946/{R}_{99}$.}
\label{M-R-chi-pos-part2}
\end{center}
\end{figure}

\begin{figure}[!h]
\begin{center}
\includegraphics[clip,scale=0.3]{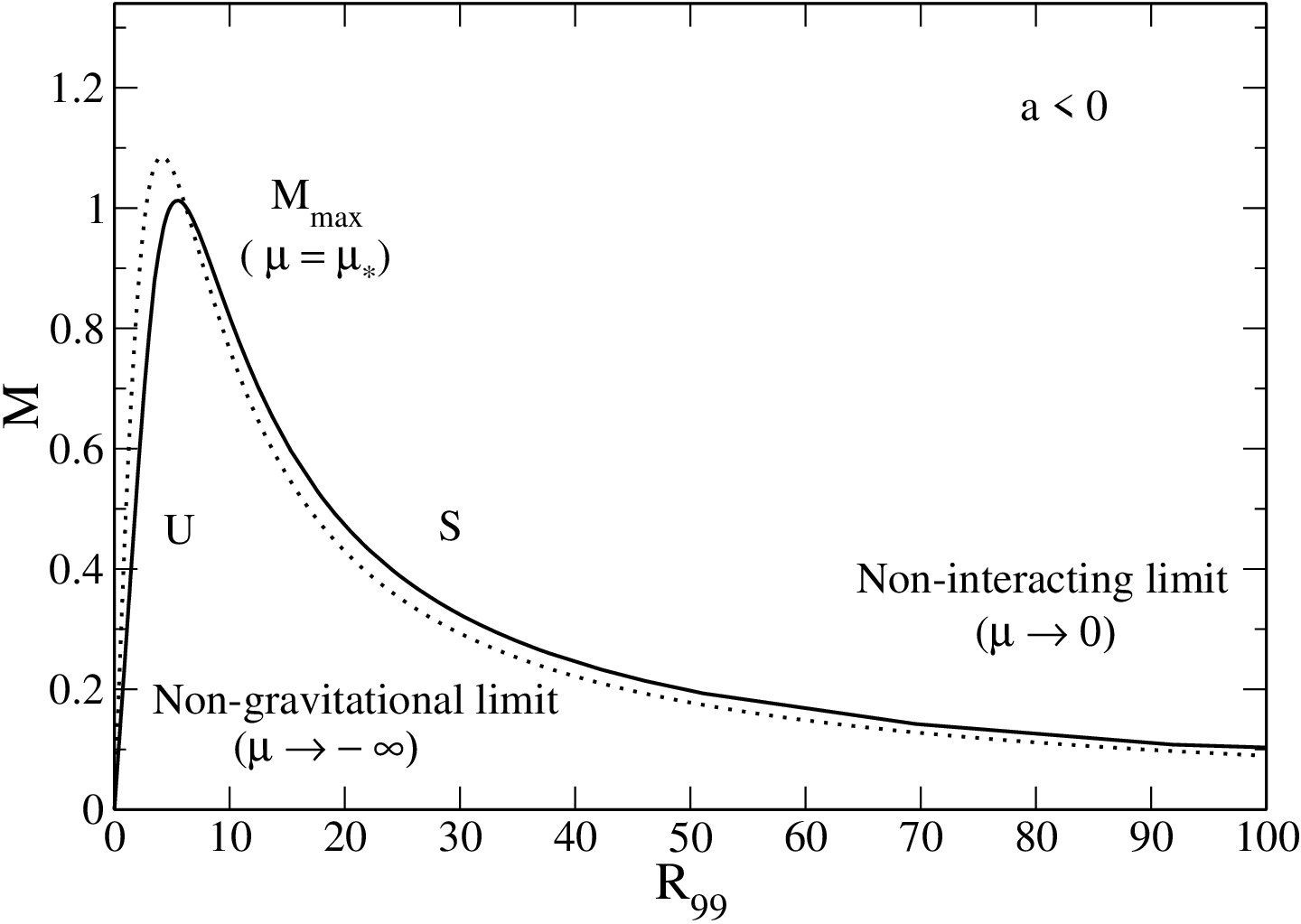}
\caption{${M}$ as a function of ${R}_{99}$ for fixed $a<0$. The mass is normalized by $M_a$ and the radius by $R_a$. In the non-interacting limit $M\rightarrow 0$ and $R_{99}\rightarrow +\infty$, we get $M\sim 9.946/R_{99}$. There exists a maximum mass $M_{max}\simeq 1.012$ corresponding to a radius $R_{99}^*\simeq 5.5$. The configurations with small radius $R<R_*$ (i.e. $\mu<\mu_{*}$ where $\mu_{*}=-1.000$ corresponds to the mass peak) are unstable. The dotted line corresponds to the approximate analytical mass-radius relation $M=2\sigma R/(\nu R^2+6\pi\zeta)$ (with $R_{99}=2.38167R$) based on the Gaussian ansatz \cite{paper1}. The radius is given as a function of the mass by $R=(\sigma/\nu M)(1\pm\sqrt{1-6\pi\zeta\nu M^2/\sigma^2})$. In the non-interacting limit $R_{99}\rightarrow +\infty$, we get $M\sim 2\sigma/\nu R$ i.e. $M^{Gauss}\sim 8.955/R_{99}$ and in the non-gravitational limit $R\rightarrow 0$ (unstable), we get $M\sim \sigma R/(3\pi\zeta)$ i.e. $M^{Gauss}\sim 0.5262 R_{99}$. On the other hand, $M_{max}^{Gauss}=\sigma/\sqrt{6\pi\zeta\nu}\simeq 1.085$ and $R_*^{Gauss}=(6\pi\zeta/\nu)^{1/2}$ i.e. $(R_{99}^*)^{Gauss}\simeq 4.125$. The analytical mass-radius relation  has the same qualitative shape as the numerical curve and provides a good quantitative agreement for any values of the mass.}
\label{M-R-chi-neg-part1}
\end{center}
\end{figure}

\begin{figure}[!h]
\begin{center}
\includegraphics[clip,scale=0.3]{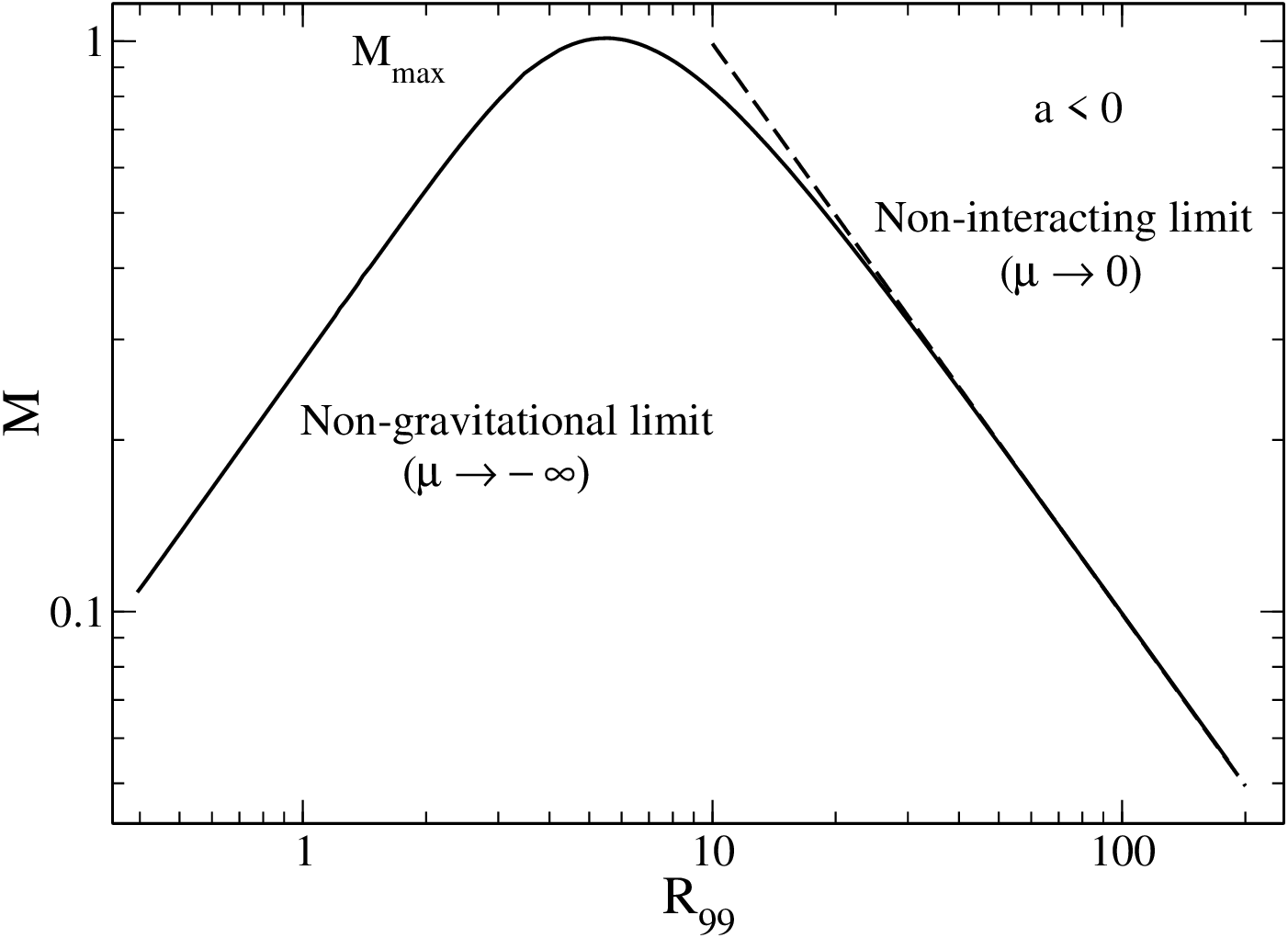}
\caption{${M}$ as a function of ${R}_{99}$ in log-log plot for fixed $a<0$. The dashed line corresponds to ${M}\sim 9.946/{R}_{99}$. }
\label{M-R-chi-neg-part2}
\end{center}
\end{figure}

The mass-radius relation for a fixed value of the scattering length is
plotted in Figs. \ref{M-R-chi-pos-part1}, \ref{M-R-chi-pos-part2},
\ref{M-R-chi-neg-part1} and \ref{M-R-chi-neg-part2} for positive and
negative scattering lengths ($R_{99}$ corresponds to the radius
containing $99\%$ of the mass).  In the non-interacting case $a=0$,
or for $M\rightarrow 0$ when $a\neq 0$, we find
\begin{equation}
\label{fa1}
R_{99}=9.946 \frac{\hbar^2}{GMm^2},
\end{equation}
in agreement with previous works \cite{rb,membrado}. In the TF limit valid for $M\rightarrow +\infty$ (when $a>0$) \cite{bohmer}, we have the analytical result
\begin{equation}
\label{fa2}
R_{99}\sim 2.998\left (\frac{a\hbar^2}{Gm^3}\right )^{1/2}.
\end{equation}
For $a>0$, there exists an equilibrium state for all the values of the mass $M$ and the configurations are stable (S), see Figs. \ref{M-R-chi-pos-part1} and \ref{M-R-chi-pos-part2}. The radius (\ref{fa2}) represents the minimum radius achievable by the system \cite{paper1}. For $a<0$, there exists a maximum mass
\begin{eqnarray}
\label{fa3}
M_{max}=1.012\frac{\hbar}{\sqrt{|a|Gm}},
\end{eqnarray}
corresponding to the radius
\begin{eqnarray}
\label{fa4}
R^*_{99}=5.5\left (\frac{|a|\hbar^2}{Gm^3}\right )^{1/2}.
\end{eqnarray}
There is no equilibrium state with $M>M_{max}$, see Figs. \ref{M-R-chi-neg-part1} and \ref{M-R-chi-neg-part2}. In that case, the system is expected to collapse and form a black hole \footnote{The maximum mass (\ref{fa3}) that we obtain in the case of negative scattering lengths (attractive short-range interactions) is a purely Newtonian result. In this sense, it is
very different from the maximum mass of white dwarf stars \cite{chandra} and
boson stars \cite{kaup,rb,colpi} whose existence is due to relativistic effects. Attractive short-range
interactions add to the gravitational attraction and make the system very
unstable. This is why there exists a maximum mass even in the Newtonian theory.
Of course, if we want to describe the collapse of the system for $M>M_{max}$ and the formation of a black
hole, we must ultimately resort to general relativity.}. For $M<M_{max}$, the right branch ($R>R_*$) is stable (S) and the left branch ($R<R_*$) is unstable (U). As indicated previously, the change of stability occurs at the turning point of mass in agreement with the Poincar\'e theorem (see Paper I and Sec. \ref{sec_poincare}).

The mass-radius relation $M(R)$ for a fixed value of the scattering length $a$ has been studied in Paper I by using a Gaussian ansatz. This leads to the approximate analytical relation
\begin{eqnarray}
\label{fa5}
M=\frac{2\sigma}{\nu}\frac{\frac{\hbar^2}{Gm^2 R}}{1-\frac{6\pi\zeta a\hbar^2}{\nu Gm^3R^2}},
\end{eqnarray}
with $\sigma=3/4$, $\zeta=1/(2\pi)^{3/2}$, $\nu=1/\sqrt{2\pi}$. We can also express the radius as a function of the mass as
\begin{eqnarray}
\label{fa5inv}
R=\frac{\sigma}{\nu}\frac{\hbar^2}{GMm^2}\left (1\pm\sqrt{1+\frac{6\pi\zeta \nu}{\sigma^2}\frac{GmM^2a}{\hbar^2}}\right ),
\end{eqnarray}
with $+$ when $a\ge 0$ and $\pm$ when $a<0$. This relation is compared with the exact mass-radius relation in Figs. \ref{M-R-chi-pos-part1} and \ref{M-R-chi-neg-part1} for positive and negative scattering lengths. The analytical model gives the same scalings as Eqs. (\ref{fa1}), (\ref{fa2}), (\ref{fa3}) and (\ref{fa4}) with the  prefactors $8.955$, $4.125$, $1.085$ and $4.125$ respectively. We see that the agreement between the analytical relation and the numerical one is qualitatively correct in all cases. It is also quantitatively good, except in the TF limit. This is because, in the TF limit, the density has a compact support that is poorly represented by a Gaussian distribution.

\subsection{The mass-energy relation}
\label{sec_faenergy}

\begin{figure}[!h]
\begin{center}
\includegraphics[clip,scale=0.3]{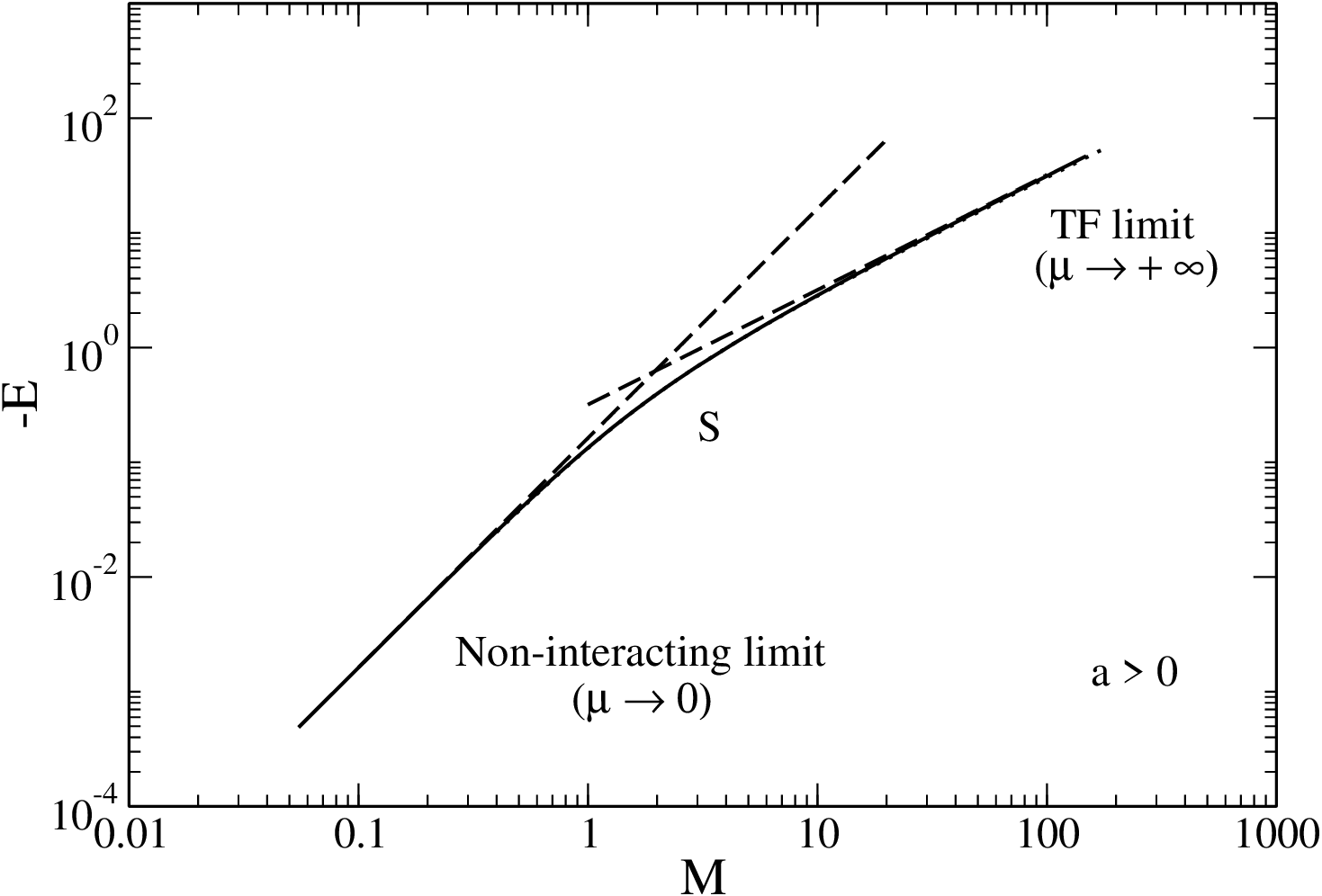}
\caption{${E}$ as a function of ${M}$ for fixed $a>0$. The eigenenergy is normalized by $E_a'$ and the mass by $M_a$. In the non-interacting limit $M\rightarrow 0$, we get ${E}\sim -0.1628 {M}^2$ and in the TF limit $M\rightarrow +\infty$, we obtain ${E}\sim -0.3183{M}$. These asymptotes are represented in dashed lines. We have also represented in dotted line the analytic expression obtained from the Gaussian ansatz. It is given by  $E=\sigma/R^2+4\pi\zeta M/R^3-2\nu M/R$ where $R$ is related to $M$ by the equation given in the caption of Fig. \ref{M-R-chi-pos-part1}. In the non-interacting limit $M\rightarrow 0$, we get $E\sim -(3\nu^2/4\sigma)M^2$ yielding $E^{Gauss}\sim -0.1592 M^2$ and in the TF limit $M\rightarrow +\infty$, we get $E\sim -(4\nu^{3/2}/(3(6\pi\zeta)^{1/2}))M$ yielding $E^{Gauss}\sim -0.3071M$. The agreement is very good for all values of mass and we hardly see the difference between the two curves.}
\label{E-M-chi-pos-new}
\end{center}
\end{figure}

\begin{figure}[!h]
\begin{center}
\includegraphics[clip,scale=0.3]{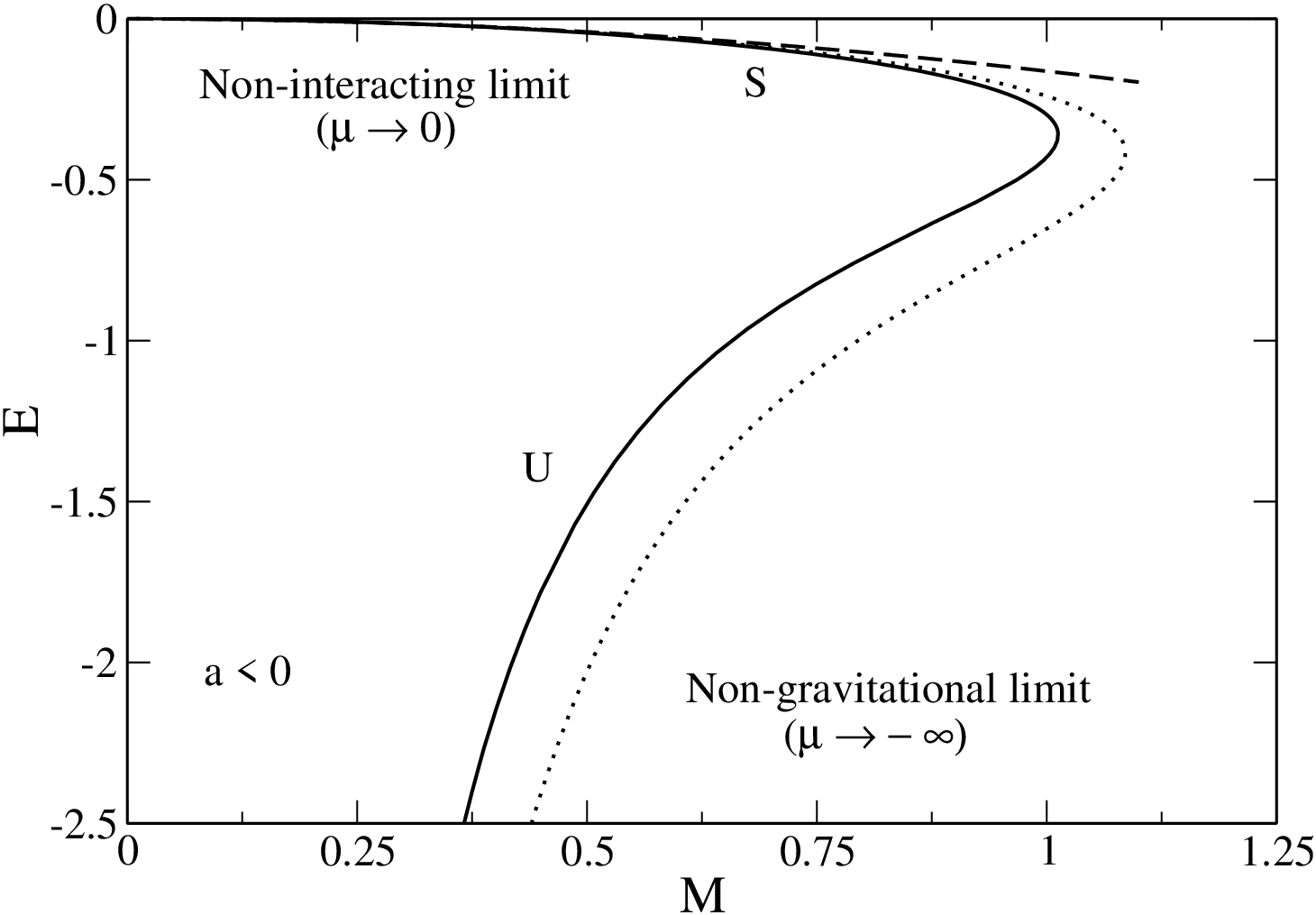}
\caption{${E}$ as a function of ${M}$ for fixed $a<0$. The eigenenergy is normalized by $E_a'$ and the mass by $M_a$. In the non-interacting limit $M\rightarrow 0$, we get ${E}\sim -0.1628 {M}^2$. The eigenenergy corresponding to the maximum mass $M_{max}\simeq 1.012$ is $E_*\simeq -0.36$. We have also represented in dotted line the analytic expression obtained from the Gaussian ansatz.  It is given by $E=\sigma/R^2-4\pi\zeta M/R^3-2\nu M/R$ where $R$ is related to $M$ by the equation given in the caption of Fig. \ref{M-R-chi-neg-part1}. In the non-interacting limit $M\rightarrow 0$ with $E\rightarrow 0$, we get $E\sim -(3\nu^2/4\sigma)M^2$ yielding $E^{Gauss}\sim -0.1592M^2$ and in the non-gravitational  limit $M\rightarrow 0$ with $E\rightarrow -\infty$, we get $E\sim -(\sigma^{3}/(3(3\pi\zeta)^{2}))M^{-2}$ yielding $E^{Gauss}\sim -0.3927/M^2$. The eigenenergy corresponding to the maximum mass $M_{max}^{Gauss}\simeq 1.085$ is $E_*^{Gauss}=-5\sigma\nu/(18\pi\zeta)\simeq -0.4166$.}
\label{E-M-chi-neg-new}
\end{center}
\end{figure}

The eigenenergy $E$ is plotted as a function of the mass $M$ in Figs. \ref{E-M-chi-pos-new} and \ref{E-M-chi-neg-new} for positive and negative scattering lengths.  In the non-interacting case $a=0$, or for  $M\rightarrow 0$ when $a\neq 0$, we obtain
\begin{eqnarray}
\label{fa6}
E=-0.1628\frac{G^2M^2m^3}{\hbar^{2}}.
\end{eqnarray}
In the TF limit valid for $M\rightarrow +\infty$ (when $a>0$), we have the analytical result \cite{paper1}:
\begin{eqnarray}
\label{fa7}
E=-\frac{1}{\pi}\frac{G^{3/2}m^{5/2}M}{a^{1/2}\hbar}.
\end{eqnarray}
For $a<0$, the eigenenergy corresponding to the point of maximum mass is
\begin{eqnarray}
\label{fa8}
E_*=-0.36\frac{Gm^2}{|a|}.
\end{eqnarray}

The eigenenergy $E$ can be estimated analytically from the Gaussian ansatz by using the results of Paper I (see in particular Sec. III.B). Using Eqs. (I-77) and (I-83), we get
\begin{eqnarray}
\label{fa9}
E=\sigma \frac{\hbar^2}{mR^2}+2{\zeta}\frac{2\pi a \hbar^2 M}{m^2R^{3}}-2\nu \frac{GMm}{R},
\end{eqnarray}
where the mass and the radius are related to each other by Eq. (\ref{fa5}). We can therefore express $E$ as a function of $R$ alone
\begin{equation}
\label{qs1}
E=-\sigma\frac{\hbar^2}{mR^2}\left (4\frac{1-\frac{2\pi\zeta a\hbar^2}{\nu Gm^3R^2}}{1-\frac{6\pi\zeta a\hbar^2}{\nu Gm^3R^2}}-1\right ).
\end{equation}
Then, eliminating the radius between $M(R)$ and $E(R)$, we  obtain $E$ as a function of $M$ in parametric form. Alternatively, we can substitute the analytical expression  of $R(M)$ given by Eq. (\ref{fa5inv}) in Eq. (\ref{fa9}).  In the non-interacting case, using $E=3W/2N$ \cite{paper1} and Eqs. (I-83) and (I-93), we obtain
\begin{eqnarray}
\label{fa10}
E=-\frac{3\nu^2}{4\sigma}\frac{G^2M^2m^3}{\hbar^{2}}.
\end{eqnarray}
In the TF approximation (for $a>0$), using $E=4W/3N$ \cite{paper1} and Eqs. (I-83) and (I-94), we find that
\begin{eqnarray}
\label{fa11}
E=-\frac{4\nu^{3/2}}{3(6\pi\zeta)^{1/2}}\frac{G^{3/2}m^{5/2}M}{a^{1/2}\hbar}.
\end{eqnarray}
In the non-gravitational limit (for $a<0$), using $E=-\Theta_Q/3N$ \cite{paper1} and Eqs. (I-83) and (I-96), we get
\begin{eqnarray}
\label{fa12}
E=-\frac{\sigma^{3}}{3(3\pi\zeta)^{2}}\frac{\hbar^{2}m}{M^2a^{2}}.
\end{eqnarray}
Finally, combining Eqs. (\ref{fa9}), (I-99) and (I-100),  the eigenenergy corresponding to the point of maximum mass is found to be
\begin{eqnarray}
\label{fa13}
E_*=-\frac{5\sigma\nu}{18\pi\zeta}\frac{Gm^2}{|a|}.
\end{eqnarray}
This returns the scalings of Eqs. (\ref{fa6}), (\ref{fa7}) and (\ref{fa8}) with the prefactors $-0.1592$, $-0.3071$ and $-0.4166$ (in the non-gravitational case, the prefactor is $-0.3927$). We see in Figs.  \ref{E-M-chi-pos-new} and \ref{E-M-chi-neg-new} that the agreement with the numerical curves is fairly good, even in the TF limit. This is because the value of the energy is less sensitive to  the detailed form of the density profile. Close to the maximum mass (for $a<0$), the agreement is good but not excellent.

\begin{figure}[!h]
\begin{center}
\includegraphics[clip,scale=0.3]{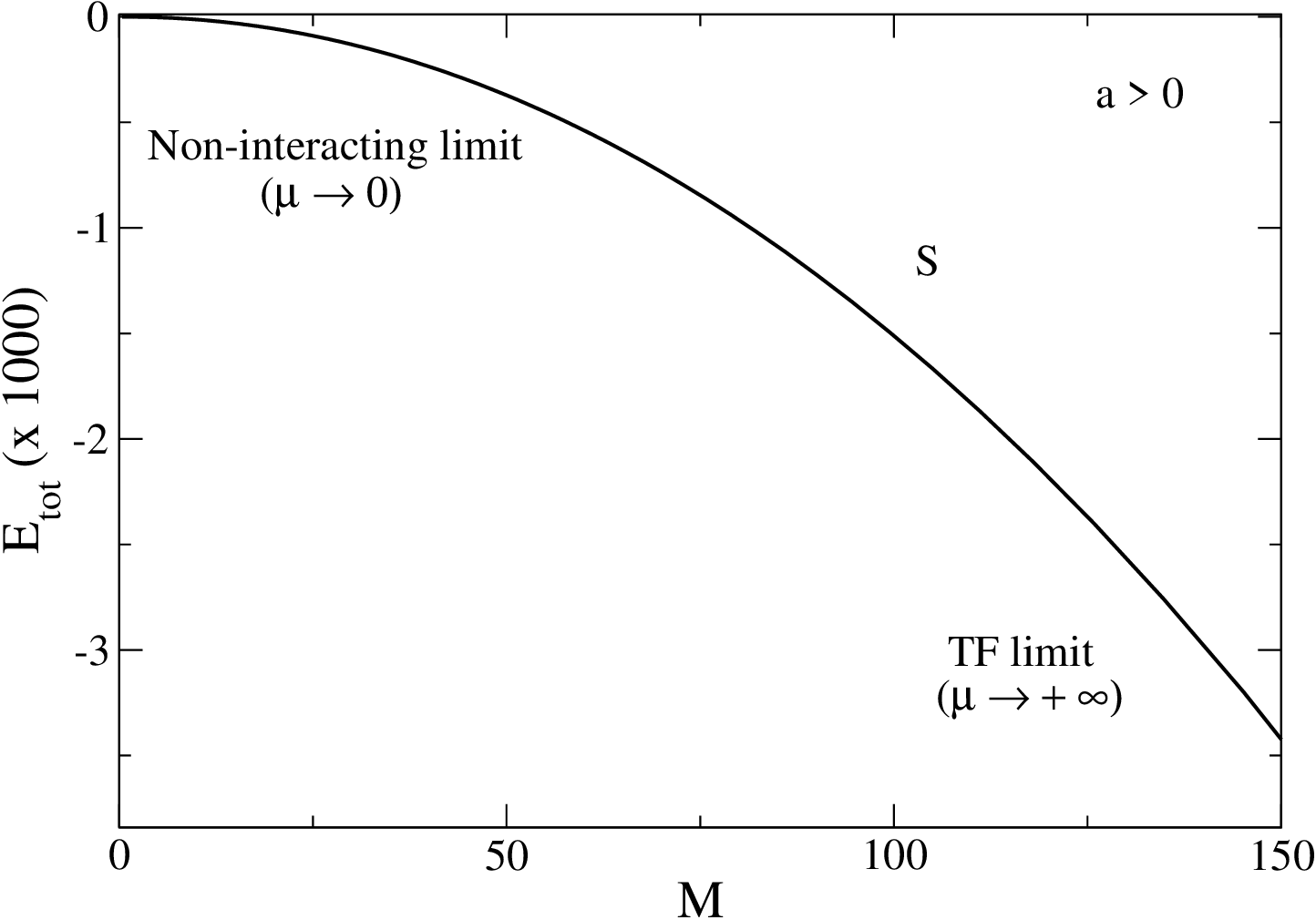}
\caption{$E_{tot}$ (obtained with the Gaussian ansatz) as a function of $M$ for fixed $a>0$. The energy is normalized by $E_a$ and the mass by $M_a$. This yields $E_{tot}=\sigma M/R^2+2\pi\zeta M^2/R^3-\nu M^2/R$ where $R$ is related to $M$ by the equation given in the caption of Fig. \ref{M-R-chi-pos-part1}. In the non-interacting limit $M\rightarrow 0$, we get $E_{tot}\sim -(\nu^2/4\sigma)M^3$ yielding $E_{tot}^{Gauss}\sim -0.05307M^3$ and in the TF limit $M\rightarrow +\infty$, we get $E_{tot}\sim -(2\nu^{3/2}/(3(6\pi\zeta)^{1/2}))M^2$ yielding $E_{tot}^{Gauss}\sim  -0.1536 M^2$.}
\label{ETOT-a-pos}
\end{center}
\end{figure}

\begin{figure}[!h]
\begin{center}
\includegraphics[clip,scale=0.3]{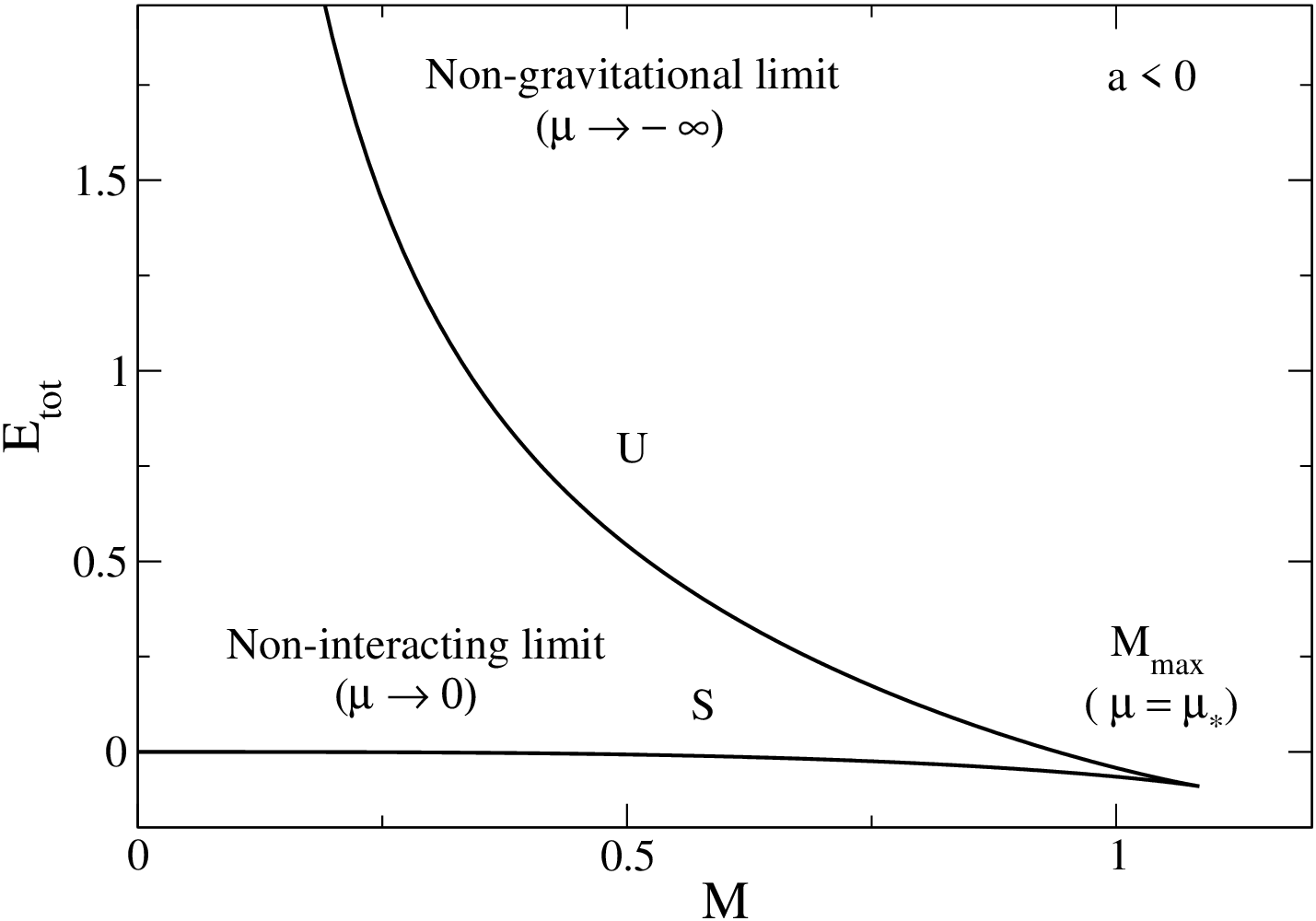}
\caption{$E_{tot}$ (obtained with the Gaussian ansatz) as a function of $M$ for fixed $a<0$. The energy is normalized by $E_a$ and the mass by $M_a$. This yields $E_{tot}=\sigma M/R^2-2\pi\zeta M^2/R^3-\nu M^2/R$ where $R$ is related to $M$ by the equation given in the caption of Fig. \ref{M-R-chi-neg-part1}. In the non-interacting limit $M\rightarrow 0$ with $E_{tot}\rightarrow 0$, we get $E_{tot}\sim -(\nu^2/4\sigma)M^3$ yielding $E_{tot}^{Gauss}\sim -0.05307M^3$ and in the non-gravitational  limit $M\rightarrow 0$ with $E_{tot}\rightarrow +\infty$, we get $E_{tot}\sim (\sigma^{3}/(3(3\pi\zeta)^{2}))M^{-1}$ yielding $E_{tot}^{Gauss}\sim 0.3927M^{-1}$. The energy corresponding to the maximum mass $M_{max}^{Gauss}\simeq 1.085$ is $E_{tot}^{Gauss}=-\sigma^{2}\nu^{1/2}/(3(6\pi\zeta)^{3/2})\simeq -0.09045$. The curve makes a spike at $M=M_{max}$ corresponding to $\mu=\mu_*$. For $M<M_{max}$, there are two solutions for the same mass but the stable state ($\mu>\mu_*$) corresponds to the state of lowest total energy $E_{tot}$, as expected (note that it corresponds to the state of highest eigenenergy $E$, i.e.  highest chemical potential $\alpha$, in Fig. \ref{E-M-chi-neg-new}).} 
\label{ETOT-a-neg}
\end{center}
\end{figure}

We can also estimate the total energy from  the Gaussian ansatz by using the results of Paper I (see in particular Sec. III.B).  Using Eqs. (I-75) and (I-83), we get
\begin{eqnarray}
\label{fa14}
E_{tot}=\sigma \frac{\hbar^2M}{m^2R^2}+{\zeta}\frac{2\pi a\hbar^2 M^{2}}{m^3R^{3}}-\nu \frac{GM^2}{R}
,
\end{eqnarray}
where the mass and the radius are related to each other by Eq. (\ref{fa5}). We can therefore express $E_{tot}$ as a function of $R$ alone
\begin{equation}
\label{qs2}
E_{tot}=-\frac{2\sigma^2}{\nu}\left (2\frac{1-\frac{2\pi\zeta a\hbar^2}{\nu Gm^3R^2}}{1-\frac{6\pi\zeta a\hbar^2}{\nu Gm^3R^2}}-1\right )\frac{\frac{\hbar^4}{Gm^4R^3}}{1-\frac{6\pi\zeta a\hbar^2}{\nu Gm^3R^2}}.
\end{equation}
Then, eliminating the radius between $M(R)$ and $E_{tot}(R)$, we  obtain $E_{tot}$ as a function of $M$ in parametric form. Alternatively, we can substitute the analytical expression  of $R(M)$ given by Eq. (\ref{fa5inv}) in Eq. (\ref{fa14}). In the non-interacting case, using $E_{tot}=NE/3$ \cite{paper1}, we obtain
\begin{eqnarray}
\label{fa15}
E_{tot}=-\frac{\nu^2}{4\sigma}\frac{G^2M^3m^2}{\hbar^{2}}.
\end{eqnarray}
In the TF approximation (for $a>0$), using $E_{tot}=NE/2$ \cite{paper1}, we find that
\begin{eqnarray}
\label{fa16}
E_{tot}=-\frac{2\nu^{3/2}}{3(6\pi\zeta)^{1/2}}\frac{G^{3/2}m^{3/2}M^2}{a^{1/2}\hbar}.
\end{eqnarray}
In the non-gravitational limit (for $a<0$), using $E_{tot}=-NE$ \cite{paper1}, we get
\begin{eqnarray}
\label{fa17}
E_{tot}=\frac{\sigma^{3}}{3(3\pi\zeta)^{2}}\frac{\hbar^{2}}{Ma^{2}}.
\end{eqnarray}
Finally, combining Eqs. (\ref{fa14}), (I-99) and (I-100), the total energy corresponding to the point of maximum mass is found to be
\begin{eqnarray}
\label{fa18}
E_{tot}^*=-\frac{\sigma^{2}\nu^{1/2}}{3(6\pi\zeta)^{3/2}}\frac{\hbar (Gm)^{1/2}}{|a|^{3/2}}.
\end{eqnarray}
The prefactors are $-0.05307$, $-0.1536$, $0.3927$ and $-0.09045$ respectively. The total energy $E_{tot}$ obtained with the Gaussian ansatz is plotted as a function of the mass $M$ in Figs. \ref{ETOT-a-pos} and \ref{ETOT-a-neg} for positive and negative scattering lengths.

\subsection{The density profiles}
\label{sec_dp}

\begin{figure}[!h]
\begin{center}
\includegraphics[clip,scale=0.3]{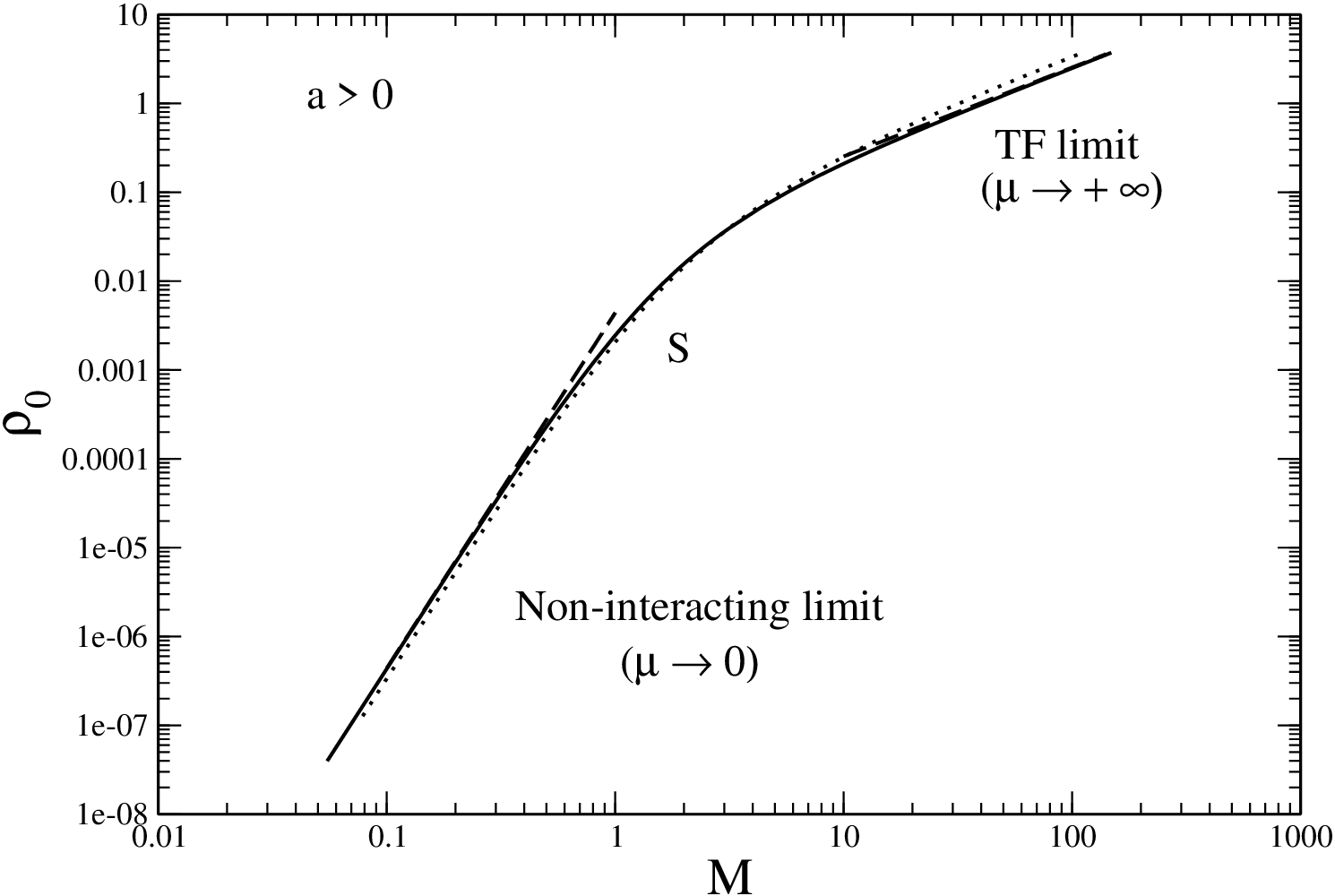}
\caption{${\rho}_0$ as a function of ${M}$ for fixed $a>0$. The central density is normalized by $\rho_a$ and the mass by $M_a$. In the non-interacting limit $M\rightarrow 0$, we get $\rho_0\sim 4.400\, 10^{-3}{M}^4$ and in the TF limit $M\rightarrow +\infty$, we get $\rho_0\sim 2.533\, 10^{-2}M$. The system is always stable. We have also represented in dotted line the analytic expression obtained from the Gaussian ansatz. It is given by  ${\rho}_0=M/\pi^{3/2}{R}^3$ where $R$ is related to $M$ by the equation given in the caption of Fig. \ref{M-R-chi-pos-part1} yielding ${\rho}_0=(\nu^3M^4/\pi^{3/2}{\sigma}^3) (1+\sqrt{1+6\pi\zeta\nu M^2/\sigma^2})^{-3}$. In the non-interacting limit $M\rightarrow 0$, we get $\rho_0\sim (\nu^3/8\sigma^3\pi^{3/2})M^4$ yielding $\rho_0^{Gauss}\sim 0.003378 M^4$ and in the TF limit $M\rightarrow +\infty$, we get ${\rho}_0=({\nu}/{6\pi^2\zeta})^{3/2}M$ yielding $\rho_0^{Gauss}\sim 0.03456M$.}
\label{rho0-M-chi-pos}
\end{center}
\end{figure}

\begin{figure}[!h]
\begin{center}
\includegraphics[clip,scale=0.3]{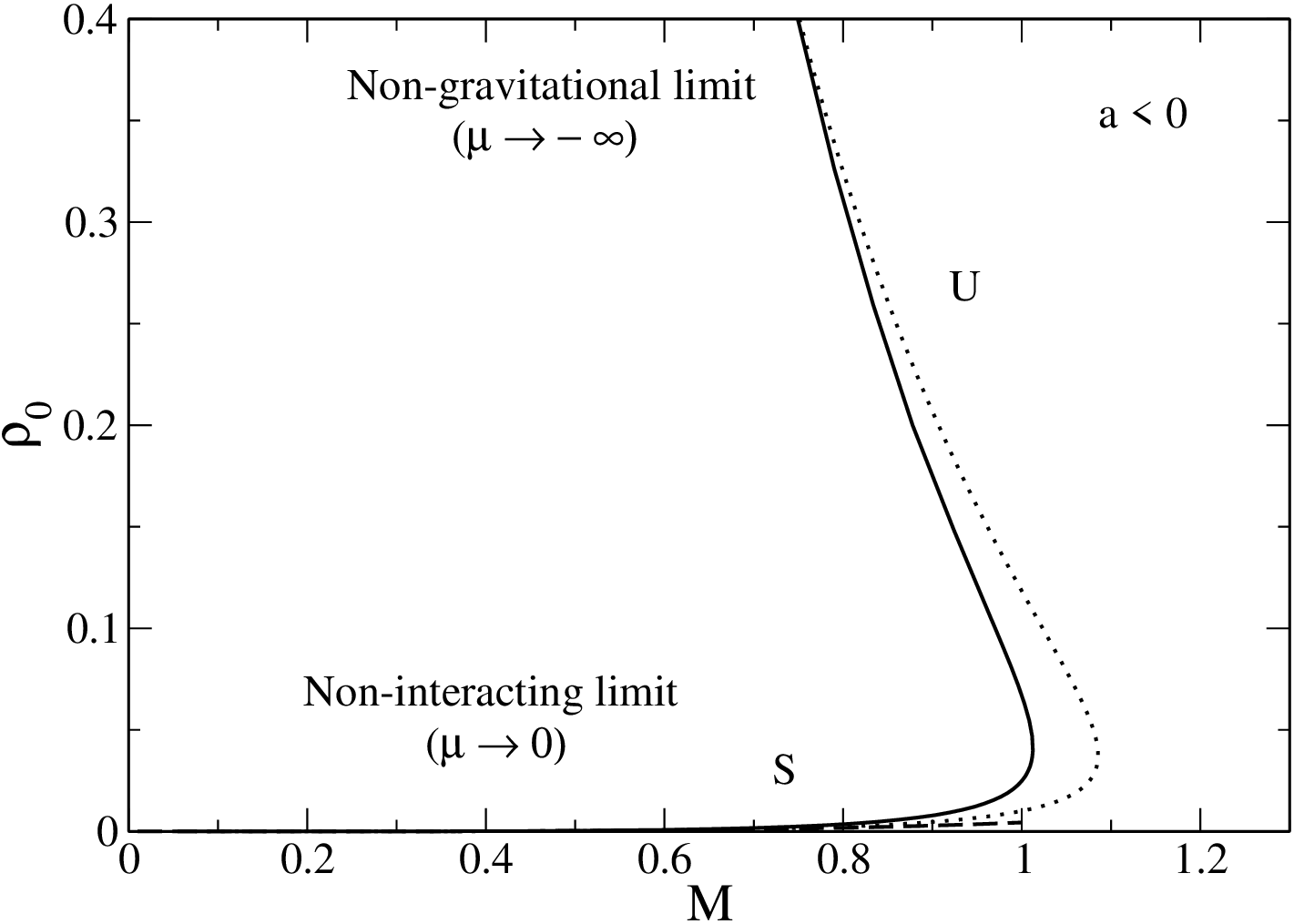}
\caption{${\rho}_0$ as a function of ${M}$ for fixed $a<0$. The central density is normalized by $\rho_a$ and the mass by $M_a$. In the non-interacting limit $M\rightarrow 0$ with $\rho_0\rightarrow 0$, we get $\rho_0\sim 4.400\, 10^{-3}{M}^4$. The central density at the point of maximum mass $M_{max}\simeq 1.012$ is $(\rho_0)_*\simeq 0.04$. The configurations with high central density $\rho_0>(\rho_0)_*$ (corresponding to $\mu<\mu_*$), located after the mass peak, are unstable. We have also represented in dotted line the analytic expression obtained from the Gaussian ansatz. It is given by  ${\rho}_0=M/\pi^{3/2}{R}^3$ where $R$ is related to $M$ by the equation given in the caption of Fig. \ref{M-R-chi-neg-part1} yielding ${\rho}_0=(\nu^3M^4/\pi^{3/2}{\sigma}^3) (1\pm\sqrt{1-6\pi\zeta\nu M^2/\sigma^2})^{-3}$. In the non-interacting limit $M\rightarrow 0$ with $\rho_0\rightarrow 0$, we get $\rho_0\sim (\nu^3/8\sigma^3\pi^{3/2})M^4$ yielding $\rho_0^{Gauss}\sim 0.003378 M^4$. In the non-gravitational limit $M\rightarrow 0$ with $\rho_0\rightarrow +\infty$, we get $\rho_0\sim (\sigma/3\pi^{3/2}\zeta)^3M^{-2}$ yielding $\rho_0^{Gauss}\sim 0.3535/M^2$. At the point of maximum mass, the central density is $(\rho_0)_*^{Gauss}=\sigma\nu/\pi^{3/2}(6\pi\zeta)^2=0.03751$}
\label{rho0-M-chi-neg}
\end{center}
\end{figure}

The central density $\rho_0$ vs mass $M$ is plotted in
Figs. \ref{rho0-M-chi-pos} and \ref{rho0-M-chi-neg} for positive and
negative scattering lengths. In the non-interacting case $a=0$, or for
$M\rightarrow 0$ when $a\neq 0$, we find that
\begin{equation}
\label{dp1}
{\rho}_0= 4.400\, 10^{-3} \frac{G^3M^4m^6}{\hbar^6}.
\end{equation}
In the TF limit valid for $M\rightarrow +\infty$ (when $a>0$) \cite{bohmer}, we have the analytical result
\begin{equation}
\label{dp2}
{\rho}_0= \frac{1}{4\pi^2}\frac{G^{3/2}Mm^{9/2}}{a^{3/2}\hbar^3}.
\end{equation}
When $a<0$, the central density at the point of maximum  mass is
\begin{equation}
\label{dp3}
({\rho}_0)_*= 0.04 \frac{Gm^4}{a^2\hbar^2}.
\end{equation}

\begin{figure}[!h]
\begin{center}
\includegraphics[clip,scale=0.3]{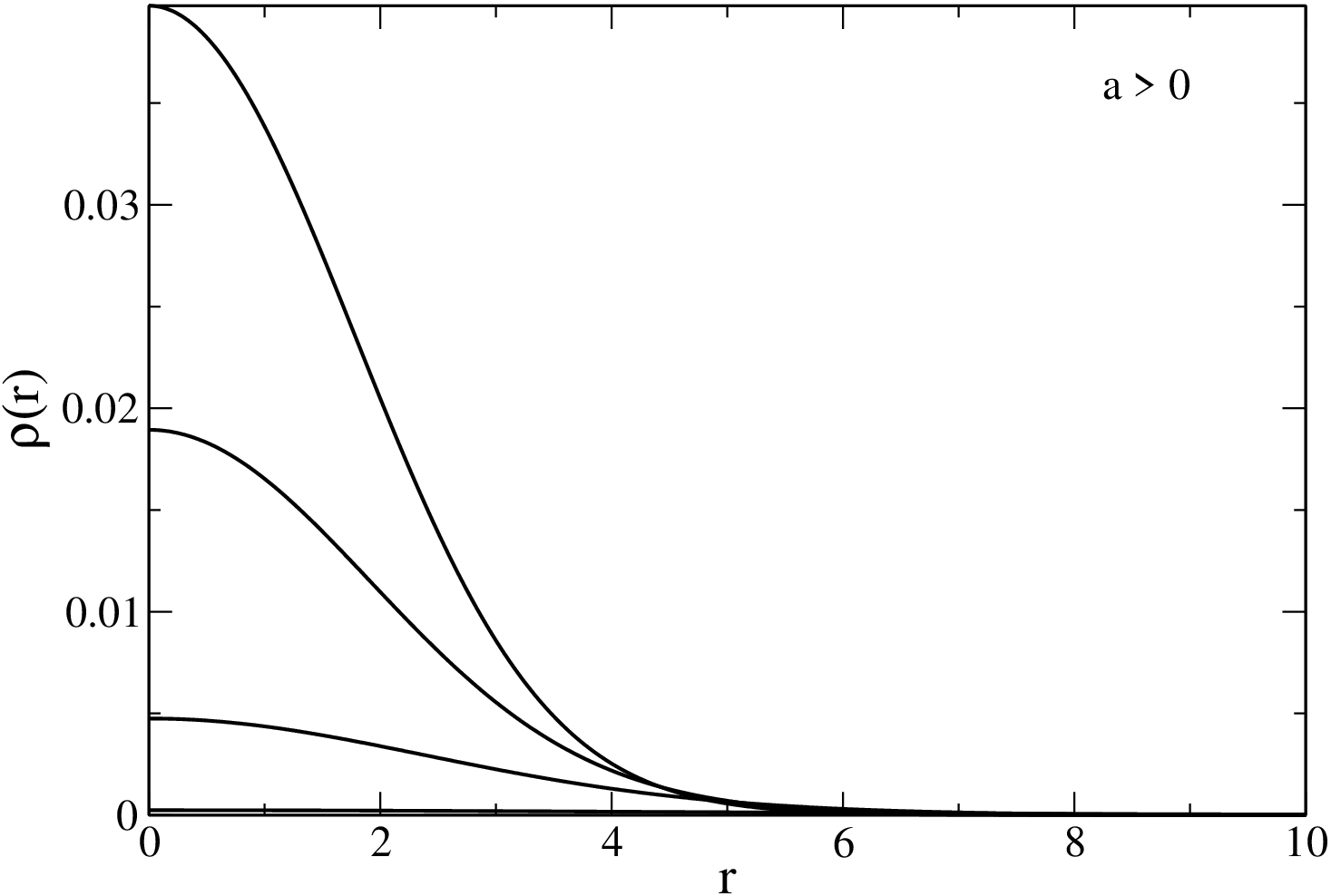}
\caption{Density profiles ${\rho}({r})$ along the series of equilibria for fixed $a>0$. The density is normalized by $\rho_a$ and the radius by $R_a$. The central density increases as $\mu$ increases. The non-interacting limit corresponds to $\mu\rightarrow 0$ and the TF limit to $\mu\rightarrow +\infty$. The different curves correspond to $\mu= 0.078, 0.34, 0.69, 1$.}
\label{rho-chi-pos-pt1}
\end{center}
\end{figure}

\begin{figure}[!h]
\begin{center}
\includegraphics[clip,scale=0.3]{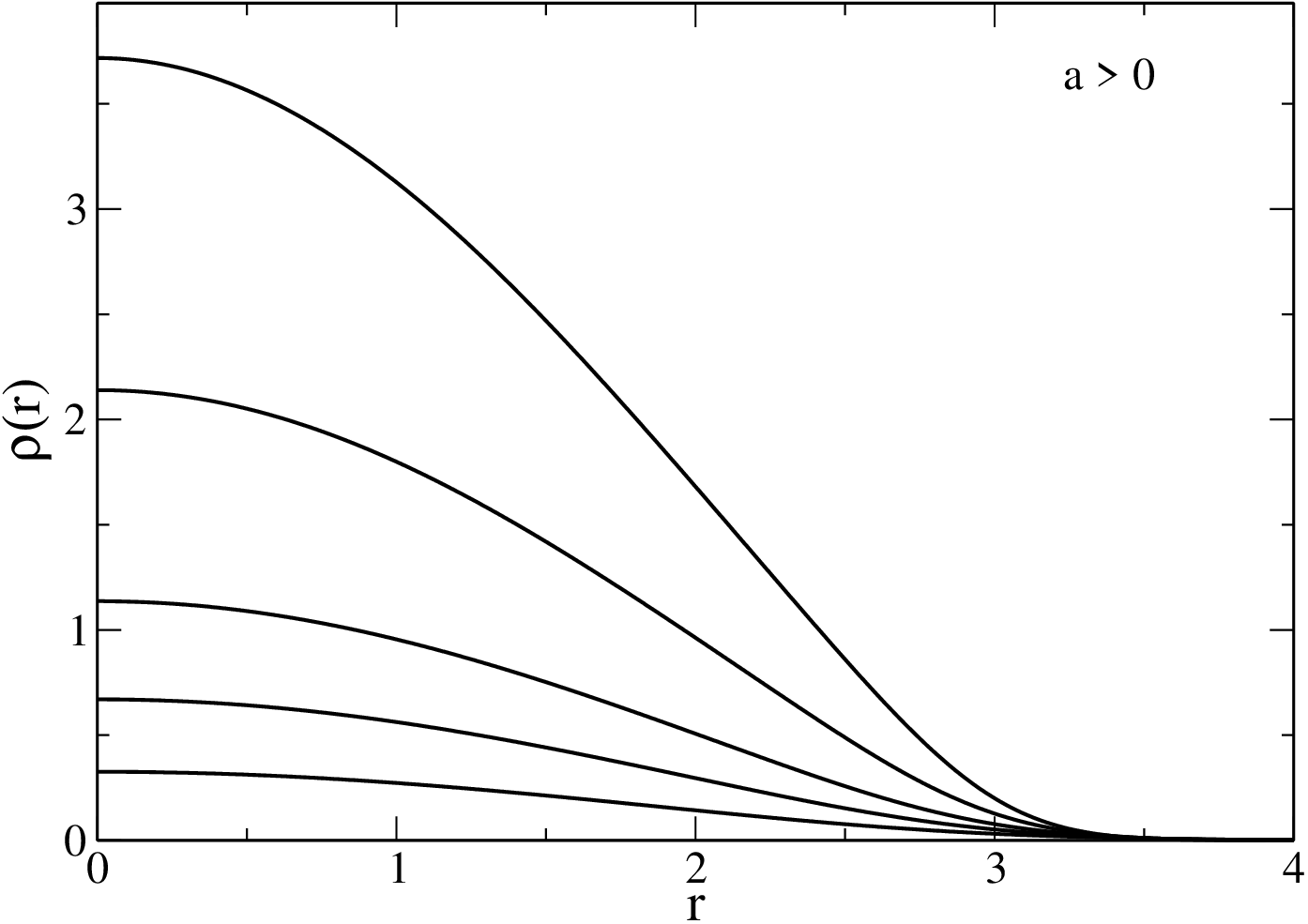}
\caption{Same as Fig. \ref{rho-chi-pos-pt1} for  $\mu= 2.86, 4.1, 5.3, 7.3, 9.66$. In the TF limit $\mu\rightarrow +\infty$, the normalized density profile tends to the asymptotic distribution $\rho(r)=(1/4\pi^2)\sin(r)/r$.}
\label{rho-chi-pos-pt2}
\end{center}
\end{figure}

\begin{figure}[!h]
\begin{center}
\includegraphics[clip,scale=0.3]{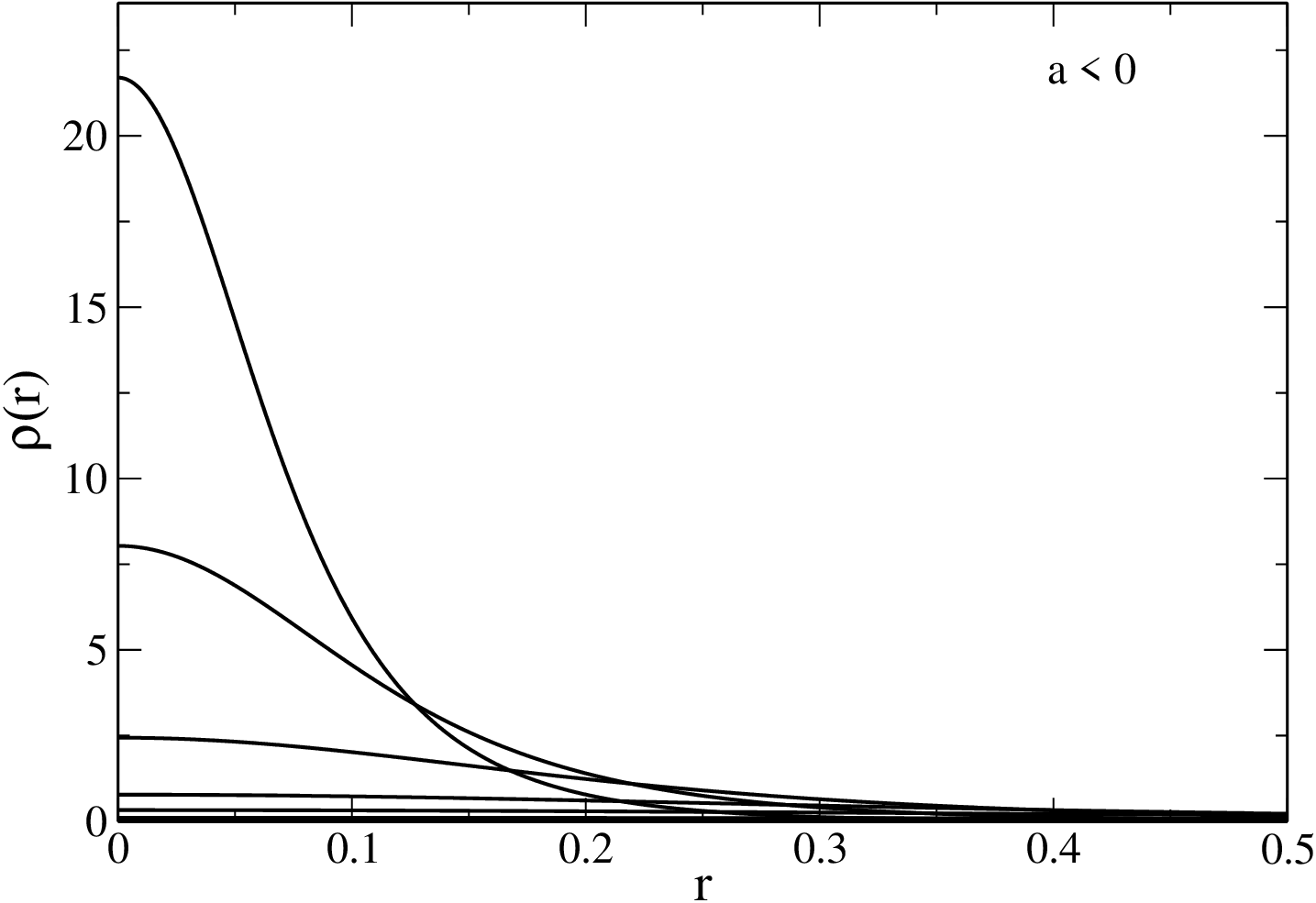}
\caption{Density profiles ${\rho}({r})$ along the series of equilibria for fixed $a<0$. The density is normalized by $\rho_a$ and the radius by $R_a$. The central density increases as $\mu$ decreases. The non-interacting limit corresponds to $\mu\rightarrow 0$ and the non-gravitational limit to $\mu\rightarrow -\infty$. The different curves correspond to $\mu=  -0.053, -0.55, -1.6, -2.86, -4.4, -7.83, -14.2, -23$. The distribution becomes unstable when $\mu<\mu_*\simeq -1.000$ corresponding to the turning point of mass (see Sec. \ref{sec_procedure}).}
\label{rho-chi-neg}
\end{center}
\end{figure}

It is interesting to compare these results with those obtained from the Gaussian ansatz. In that case,  the central density is related to the mass and to the radius by
\begin{eqnarray}
\label{dp4}
\rho_0=\frac{M}{\pi^{3/2}R^3}.
\end{eqnarray}
Combining this relation with Eq. (\ref{fa5}), we obtain the central density as a function of the radius
\begin{eqnarray}
\label{raz}
\rho_0=\frac{2\sigma}{\nu\pi^{3/2}}\frac{\frac{\hbar^2}{Gm^2 R^4}}{1-\frac{6\pi\zeta a\hbar^2}{\nu Gm^3R^2}}.
\end{eqnarray}
Eliminating the radius between $M(R)$ and $\rho_0(R)$, we obtain the approximate central density versus mass relation in parametric form. Alternatively, we can substitute the analytical expression  of $R(M)$ given by Eq. (\ref{fa5inv}) in Eq. (\ref{dp4}). In the non-interacting case, we obtain
\begin{equation}
\label{dp6}
{\rho}_0= \frac{\nu^3}{8\sigma^3\pi^{3/2}} \frac{G^3M^4m^6}{\hbar^6}.
\end{equation}
In the Thomas-Fermi limit (when $a>0$), we find that
\begin{equation}
\label{dp7}
{\rho}_0=\left (\frac{\nu}{6\pi^2\zeta}\right )^{3/2} \frac{G^{3/2}Mm^{9/2}}{a^{3/2}\hbar^3}.
\end{equation}
In the non-gravitational limit (when $a<0$), we get
\begin{equation}
\label{adp1}
{\rho}_0=\left (\frac{\sigma}{3\pi^{3/2}\zeta}\right )^{3} \frac{m^{3}}{M^2|a|^{3}}.
\end{equation}
Finally, the central density at the point of maximum mass is found to be
\begin{equation}
\label{adp2}
({\rho}_0)_*=\frac{\sigma\nu}{\pi^{3/2}(6\pi\zeta)^2} \frac{Gm^{4}}{a^{2}\hbar^2}.
\end{equation}
This returns the scalings of Eqs. (\ref{dp1}), (\ref{dp2}) and
(\ref{dp3}) with the prefactors $0.003378$, $0.03456$ and $0.03751$
(the prefactor in the non-gravitational limit is
$0.3535$). The analytical mass-central density relation is compared with the
exact relation  in Figs. \ref{rho0-M-chi-pos} and \ref{rho0-M-chi-neg}.
The agreement is fairly good, except in the TF
limit (when $a>0$). Close to the maximum mass (when $a<0$), the agreement is good but not perfect.
Some density profiles $\rho(r)$ are represented in
Figs. \ref{rho-chi-pos-pt1}, \ref{rho-chi-pos-pt2}, \ref{rho-chi-neg}
for positive and negative scattering lengths and different values of $M$.

\subsection{Stability analysis and Poincar\'e theorem}
\label{sec_poincare}

It is interesting to develop an analogy with thermodynamics. We have seen in Paper I that stable steady states of the GPP system, or equivalently of the quantum barotropic Euler-Poisson system, correspond to (local) minima of the energy functional (see Eqs. (I-56)-(I-64)):
\begin{eqnarray}
\label{enefunct}
E_{tot}[\rho,{\bf u}]=\int\rho \frac{{\bf u}^2}{2}\, d{\bf r}+\frac{\hbar^2}{2m^2}\int (\nabla\sqrt{\rho})^2\, d{\bf r}\nonumber\\
+\frac{2\pi a\hbar^2}{m^3}\int \rho^2 \, d{\bf r}+\frac{1}{2}\int\rho\Phi\, d{\bf r}.
\end{eqnarray}
at fixed mass $M$. We thus have to study the variational problem
\begin{eqnarray}
\label{vp}
\min_{\rho,{\bf u}}\lbrace E_{tot}[\rho,{\bf u}]  \, |\, M[\rho]=M\rbrace.
\end{eqnarray}
The critical points, cancelling the first order variations of constrained total energy, are given by $\delta E_{tot}-\alpha\delta M=0$ where $\alpha$ is a Lagrange multiplier associated with the conservation of mass that can be interpreted as a chemical potential. These first order variations lead to the steady state equation (\ref{s3}) provided that we make the identification $\alpha=E/m$. This shows that the eigenenergy $E$ can be interpreted as a chemical potential. As a result, Figs. \ref{E-M-chi-pos-new} and \ref{E-M-chi-neg-new} give the chemical potential $\alpha=\partial E_{tot}/\partial M$ (conjugate quantity) as a function of the mass $M$ (conserved quantity). Similarly, in thermodynamics, the statistical equilibrium state of a system is obtained by maximizing the entropy $S$ at fixed energy $E$ (see, e.g. \cite{ijmpb}). The first order variations are given by $\delta S-\beta\delta E=0$ where $\beta$ is a Lagrange multiplier associated with the conservation of energy that represents the inverse temperature. The caloric curve $\beta(E)$ gives the inverse temperature $\beta=\partial S/\partial E$ (conjugate quantity) as a function of energy $E$ (conserved quantity). Now, using the Poincar\'e theory of linear series of equilibria, we know that  when we plot the conjugate quantity as a function of the conserved quantity, a change of stability can only occur at a turning point of the conserved quantity or at a bifurcation point. In the present case, when $a<0$, the change of stability occurs at the {\it turning point} of mass (see Fig. \ref{E-M-chi-neg-new}). Similarly, in the thermodynamics of self-gravitating systems, the change of stability occurs at the turning point of energy in the microcanonical ensemble or at the turning point of temperature (equivalent to the turning point of mass) in the canonical ensemble \cite{ijmpb}. Finally, since $\delta E_{tot}=\delta M=0$ at the turning point of mass, we conclude that the curve $E_{tot}(M)$ presents a {\it cusp} at that point (see Fig. \ref{ETOT-a-neg}). Similarly, in thermodynamics, since $\delta S=\delta E=0$ at the turning point of energy,  the curve $S(E)$ presents a cusp at that point (see Fig. 4 in \cite{pre}).

Let us recapitulate:  For $a>0$, the series of equilibria containing all the critical points of the variational problem (\ref{vp}) is parametrized by $\mu$ going from $\mu\rightarrow +\infty$ (TF limit) to $\mu\rightarrow 0$ (non-interacting limit). The series of equilibria $R(M)$, $E(M)$ or $E_{tot}(M)$ is monotonic. Since the system is stable in the TF limit (see Sec. \ref{sec_procedure}), we conclude from the Poincar\'e theorem that the whole series of equilibria is stable for $a>0$. By continuity, the non-interacting BEC ($\mu=a=0$) is also stable. For $a<0$, the series of equilibria is parametrized by $\mu$ going from  $\mu\rightarrow 0$ (non-interacting limit) to $\mu\rightarrow -\infty$ (non-gravitational limit). The series of equilibria $R(M)$, $E(M)$ or $E_{tot}(M)$ is non-monotonic. There is a turning point of mass at $M=M_{max}$ corresponding to $\mu=\mu_*\simeq -1.000$. For $M>M_{max}$ there is no solution to the variational problem (\ref{vp}) and for $M<M_{max}$ there are two solutions with the same mass.  Since the system is stable in the non-interacting limit $\mu=a=0$ (as we have just seen), we conclude  from the Poincar\'e theorem that the system is stable for $\mu>\mu_*$ and unstable for $\mu<\mu_*$ (this corresponds to high central densities). This leads to the stability/instability regions shown in Figs \ref{M-R-chi-pos-part1}-\ref{rho0-M-chi-neg}.

\section{The fundamental curves for a fixed value of the mass}
\label{sec_funmassfixed}

\subsection{The radius vs scattering length relation}
\label{sec_ra}

\begin{figure}[!h]
\begin{center}
\includegraphics[clip,scale=0.3]{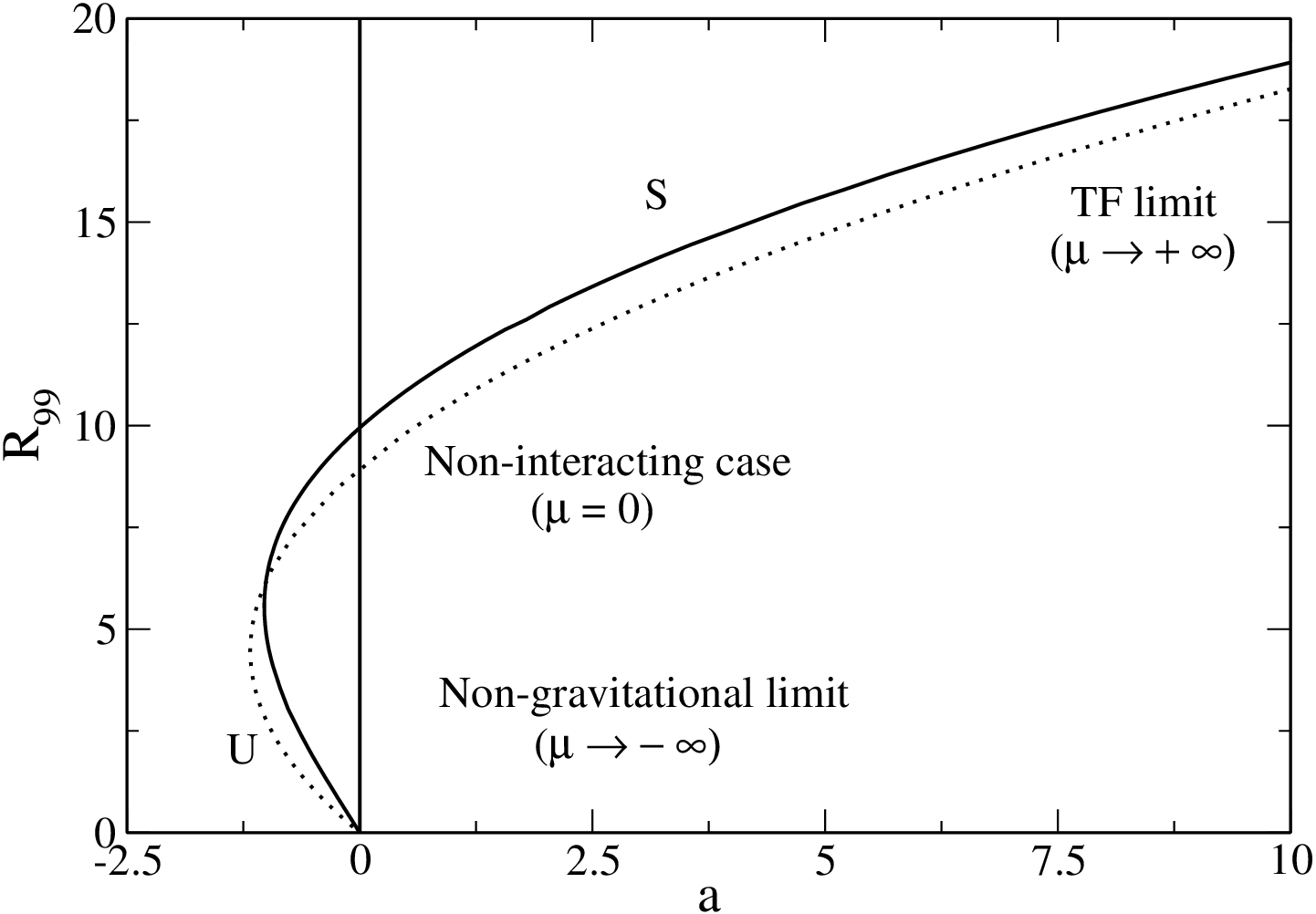}
\caption{$R_{99}$ as a function of $a$  for a fixed value of the mass $M$. The radius is normalized by $R_Q$ and the scattering length by $a_Q$. Stable solutions exist for $a\ge a_{min}\simeq -1.025$ and their radius $R_{99}\ge R_{99}^*\simeq 5.6$ is monotonically increasing with $a$. In the non-interacting case $a=0$, $R_{99}\simeq 9.946$ and in the TF limit $a\rightarrow +\infty$, $R_{99}\sim 2.998a^{1/2}$. The non-gravitational limit corresponds to $a\rightarrow 0$ and $R_{99}\rightarrow 0$ but these solutions are unstable.    The dotted line corresponds to the approximate analytical radius vs scattering length relation $a=(\nu R^2-2\sigma R)/6\pi\zeta$ (with $R_{99}=2.38167R$)  based on the Gaussian ansatz \cite{paper1}. The radius can be expressed in terms of the scattering length as $R=(\sigma/\nu)(1\pm \sqrt{1+6\pi\zeta\nu a/\sigma^2})$. In the non-interacting case $a=0$, we get $R\sim 2\sigma/\nu$ i.e. $R_{99}^{Gauss}\simeq 8.955$ and in the TF limit $a\rightarrow +\infty$, we get $R\sim (6\pi\zeta/\nu)^{1/2}a^{1/2}$ i.e. $R_{99}^{Gauss}\sim 4.125a^{1/2}$.
In the non-gravitational limit $R\rightarrow 0$, we get $R\sim (3\pi\zeta/\sigma)|a|$ i.e. $R_{99}^{Gauss}\sim 1.900 |a|$. On the other hand $a_{min}^{Gauss}=-\sigma^2/6\pi\zeta\nu\simeq -1.178$ and $R_*=\sigma/\nu$ i.e. $(R_{99}^*)^{Gauss}\simeq 4.477$. The analytical radius versus scattering length relation  has the same qualitative shape as the numerical curve and provides a good quantitative agreement except in the TF limit where the density profile differs sensibly from a Gaussian. }
\label{R99-part1}
\end{center}
\end{figure}

\begin{figure}[!h]
\begin{center}
\includegraphics[clip,scale=0.3]{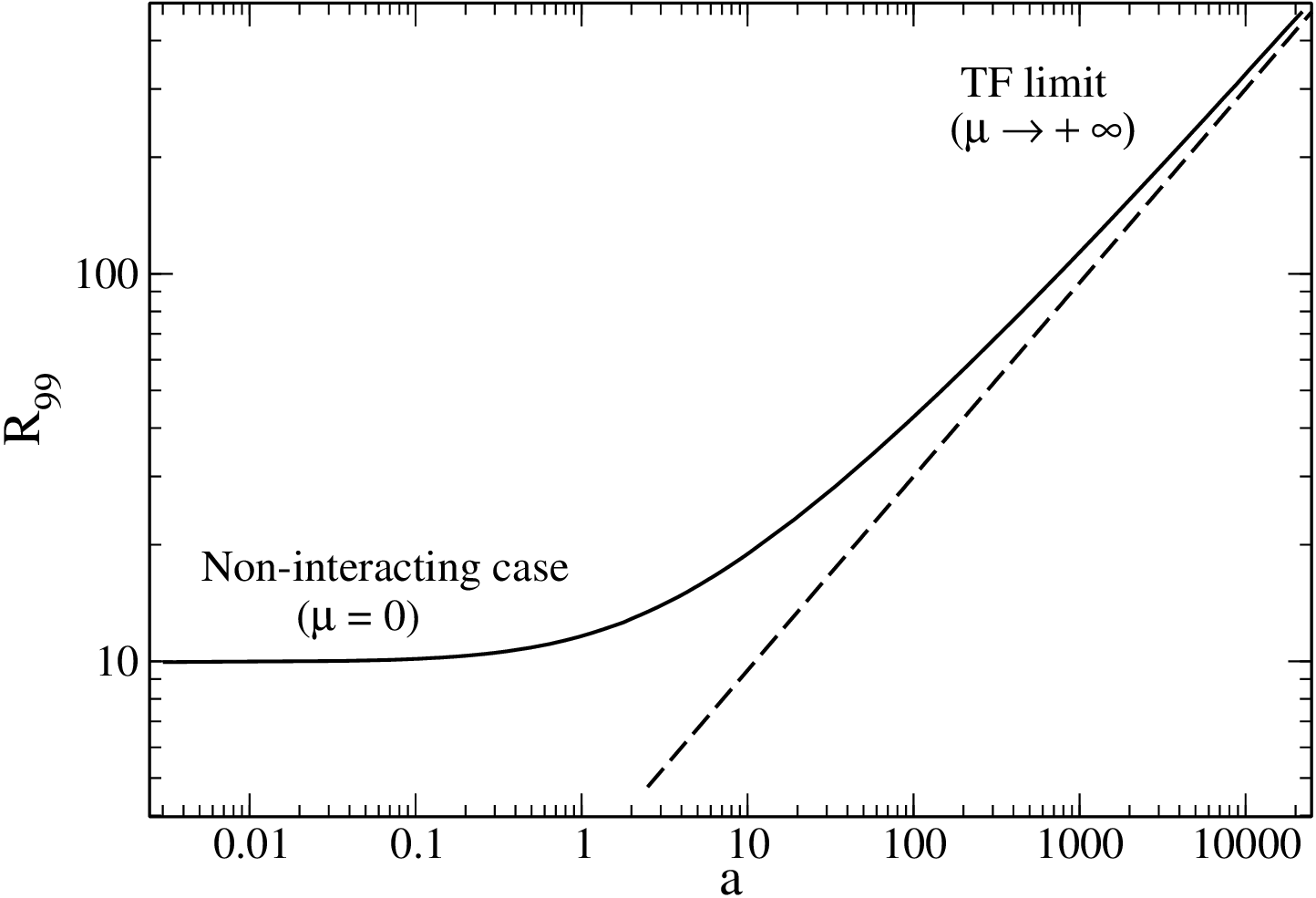}
\caption{$R_{99}$ as a function of $a$ in log-log plot for a fixed value of the mass $M$. The dashed line corresponds to $R_{99}\sim 2.998a^{1/2}$ valid in the TF limit.}
\label{R99-part2}
\end{center}
\end{figure}

The radius vs scattering length relation for a fixed value of the total mass is plotted  in Figs. \ref{R99-part1} and  \ref{R99-part2} ($R_{99}$ corresponds to the radius containing $99\%$ of the mass). In the non-interacting case $a=0$, the radius is given by Eq. (\ref{fa1}) and in the TF limit $a\rightarrow +\infty$, it is given by Eq. (\ref{fa2}). For $a<0$, there exists a minimum scattering length
\begin{eqnarray}
\label{ra1}
a_{min}=-1.025\frac{\hbar^2}{GM^2m},
\end{eqnarray}
corresponding to the radius
\begin{eqnarray}
\label{ra2}
R_{99}^*=5.6\frac{\hbar^2}{GMm^2}.
\end{eqnarray}
There is no equilibrium state when $a<a_{min}$. In that case, the system is expected to collapse and form a black hole. For $a>a_{min}$, the upper line ($R>R_*$) is stable (S) and the lower line ($R<R_*$) is unstable (U).

The radius vs scattering length relation $R(a)$ for a fixed value of the mass $M$ has been studied in Paper I by using the Gaussian ansatz. This leads to the approximate analytical relation (\ref{fa5inv}) or inversely
\begin{equation}
\label{ra3}
a=\frac{m^3}{6\pi\zeta M\hbar^2}\left (\nu GMR^2-2\sigma\frac{\hbar^2}{m^2}R\right ).
\end{equation}
The analytical model gives the same scalings as Eqs. (\ref{fa1}), (\ref{fa2}), (\ref{ra1}) and (\ref{ra2}) with the prefactors $8.955$, $4.125$, $1.178$ and $4.477$  respectively.

\subsection{The energy vs scattering length relation}
\label{sec_ea}

\begin{figure}[!h]
\begin{center}
\includegraphics[clip,scale=0.3]{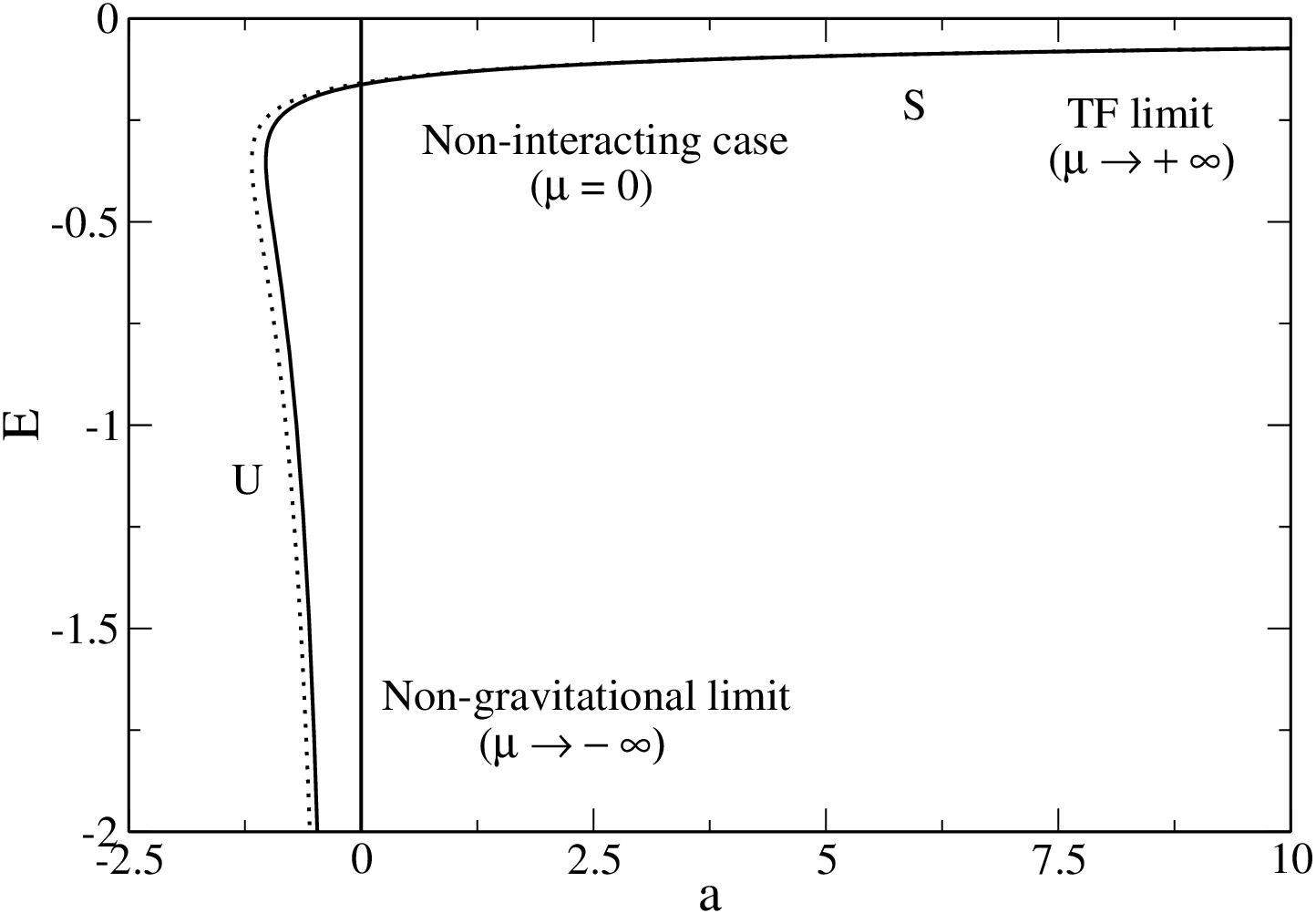}
\caption{$E$ as a function of $a$ for a fixed value of the mass $M$. The eigenenergy is normalized by $E'_Q$ and the scattering length by $a_Q$. In the non-interacting case $a=0$, we have $E\simeq -0.1628$  and in the TF limit  $a\rightarrow +\infty$, we get $E\sim -0.3183/a^{1/2}$. The eigenenergy at the point of minimum scattering length $a_{min}\simeq -1.025$ is $E_*\simeq -0.35$. We have also represented in dotted line the analytic expression obtained from the Gaussian ansatz. It is given by  $E=\sigma/R^2+4\pi\zeta a/R^3-2\nu/R$ where $R$ is related to $a$ by the equation given in the caption of Fig. \ref{R99-part1}. This yields $E=-(\sigma+4\nu R)/3R^2$ or, equivalently, $E=-(\nu^2/3\sigma)(5\pm 4\sqrt{1+6\pi\zeta\nu a/\sigma^2})/(1\pm \sqrt{1+6\pi\zeta\nu a/\sigma^2})^2$. In the non-interacting case $a=0$, we get $E\sim -(3\nu^2/4\sigma)$ yielding $E^{Gauss}\simeq -0.1592$ and in the TF limit $a\rightarrow +\infty$, we get $E\sim -(4\nu^{3/2}/(3(6\pi\zeta)^{1/2}))a^{-1/2}$ yielding $E^{Gauss}\sim -0.3071/a^{1/2}$.  In the non-gravitational  limit $M\rightarrow 0$ with $R\rightarrow 0$, we get $E\sim -(\sigma^{3}/(3(3\pi\zeta)^{2}))a^{-2}$ yielding $E^{Gauss}\sim -0.3927/a^2$. The eigenenergy corresponding to the minimum scattering length $a_{min}^{Gauss}\simeq -1.178$ is $E_*^{Gauss}=-{5\nu^2}/{3\sigma}\simeq -0.3537$. The analytical eigenenergy versus scattering length relation  has the same qualitative shape as the numerical curve and provides a good quantitative agreement for all values of the scattering length $a$. }
\label{epsilon-part1}
\end{center}
\end{figure}

\begin{figure}[!h]
\begin{center}
\includegraphics[clip,scale=0.3]{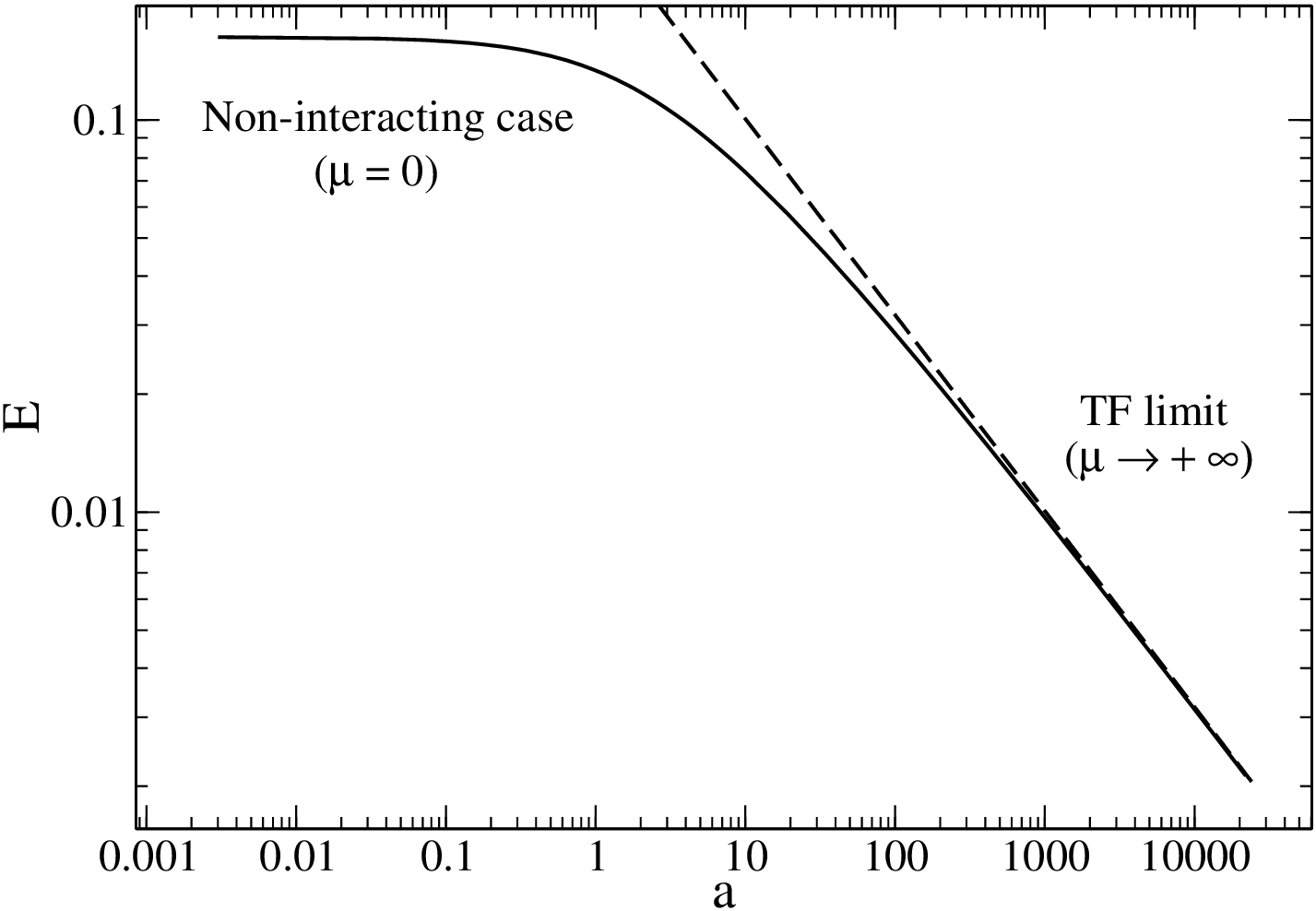}
\caption{Eigenenergy $E$ vs scattering length $a$ relation (for $a>0$) in log-log plot for a fixed value of the mass $M$. The dashed line corresponds to $E\sim -0.3183/a^{1/2}$ characterizing  the TF limit.}
\label{epsilon-part2}
\end{center}
\end{figure}

The eigenenergy $E$ vs scattering length relation for a fixed value of the total mass is plotted in Figs. \ref{epsilon-part1} and \ref{epsilon-part2}. In the non-interacting case $a=0$, the eigenenergy is given by Eq. (\ref{fa6}) and in the TF limit $a\rightarrow +\infty$, it is given by Eq. (\ref{fa7}). For $a<0$, the eigenenergy corresponding to the minimum scattering length is
\begin{eqnarray}
\label{ea1}
E_*=-0.35\frac{G^2M^2m^3}{\hbar^2}.
\end{eqnarray}

\begin{figure}[!h]
\begin{center}
\includegraphics[clip,scale=0.3]{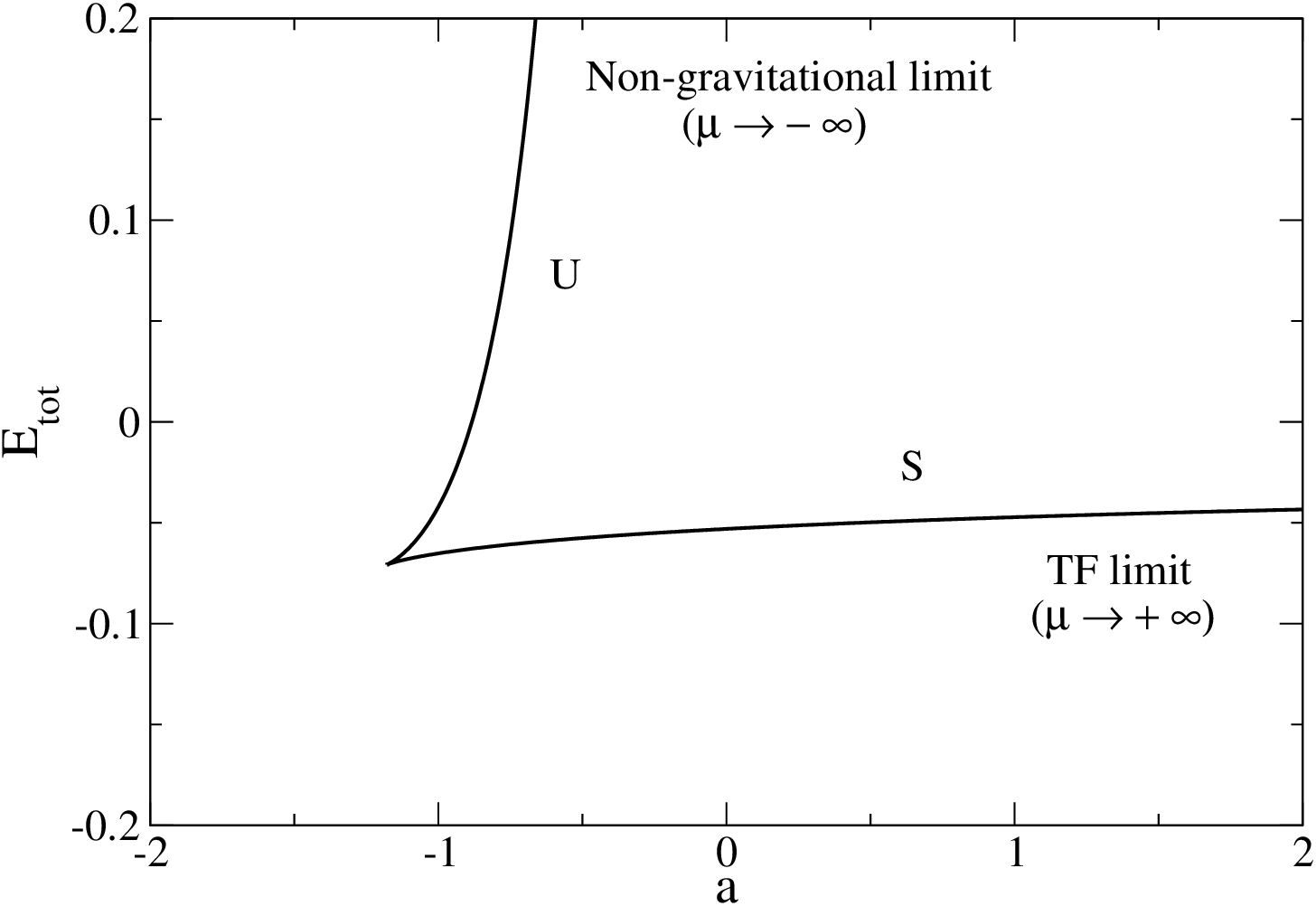}
\caption{$E_{tot}$ (obtained from the Gaussian ansatz) as a function of $a$ for a fixed value of the mass $M$. The total energy is normalized by $E_Q$ and  the scattering length by $a_Q$. This yields $E_{tot}=\sigma/R^2+ 2\pi\zeta a /R^3-\nu/R$ where $R$ is related to $a$ by the equation given in the caption of Fig. \ref{R99-part1}. This yields $E_{tot}=(\sigma-2\nu R)/3R^2$ or, equivalently, $E_{tot}=-(\nu^2/3\sigma)(1\pm 2\sqrt{1+6\pi\zeta\nu a/\sigma^2})/(1\pm \sqrt{1+6\pi\zeta\nu a/\sigma^2})^2$. In the non-interacting case $a=0$, we get $E_{tot}= -\nu^2/(4\sigma)\simeq -0.05305$ and in the TF limit $a\rightarrow +\infty$, we get $E_{tot}\sim -(2\nu^{3/2}/(3(6\pi\zeta)^{1/2}))a^{-1/2}\simeq -0.1536/\sqrt{a}$. In the non-gravitational limit $a\rightarrow 0$, we get $E_{tot}\sim (\sigma^3/(3(3\pi\zeta)^2))a^{-2}\simeq 0.3927/a^2$. The total energy at the point of minimum scattering length is $E_{tot}=-\nu^2/3\sigma\simeq -0.07074$.}
\label{ETOT-Mfixe}
\end{center}
\end{figure}

Using the Gaussian ansatz, the eigenenergy $E$ is given by Eq. (\ref{fa9}) where the radius and the scattering lengths are related to each other by Eq. (\ref{ra3}). We can therefore express $E$ as a function of $R$ alone
\begin{eqnarray}
\label{satn1}
E=-\left (\frac{\sigma \hbar^2}{m}+4\nu GMmR\right )\frac{1}{3R^2}.
\end{eqnarray}
Then, eliminating the radius between $a(R)$ and $E(R)$ we obtain $E$ as a function of $a$ in parametric form. Alternatively, we can substitute the analytical expression  of $R(a)$ given by Eq. (\ref{fa5inv}) in Eq. (\ref{fa9}). In the non-interacting case, we obtain Eq. (\ref{fa10}), in the TF limit we obtain Eq. (\ref{fa11}) and in the non-gravitational limit, we obtain Eq. (\ref{fa12}). At the minimum scattering length, we find that
\begin{eqnarray}
\label{fdfe}
E_*=-\frac{5\nu^2}{3\sigma}\frac{G^2M^2m^3}{\hbar^2}.
\end{eqnarray}
This returns the scalings of Eqs. (\ref{fa6}), (\ref{fa7}) and (\ref{ea1}) with the prefactors $-0.1592$, $-0.3071$ and  $-0.3537$ (in the non-gravitational case, the prefactor is $-0.3927$). We see in Fig. \ref{epsilon-part1} that the agreement is fairly good, even in the TF limit.

We can also use the Gaussian ansatz to estimate the total energy $E_{tot}$. It is given by Eq. (\ref{fa14})
where the radius and the scattering lengths are related to each other by Eq. (\ref{ra3}). We can therefore express $E_{tot}$ as a function of $R$ alone
\begin{eqnarray}
\label{satn2}
E_{tot}=\left (\frac{\sigma \hbar^2 M}{m^2}-2\nu GM^2R\right )\frac{1}{3R^2}.
\end{eqnarray}
Then, eliminating the radius between $a(R)$ and $E_{tot}(R)$ we obtain $E_{tot}$ as a function of $a$ in parametric form. Alternatively, we can substitute the analytical expression  of $R(a)$ given by Eq. (\ref{fa5inv}) in Eq. (\ref{fa14}). In the non-interacting case, we obtain Eq. (\ref{fa15}), in the TF limit we obtain Eq. (\ref{fa16}) and in the non-gravitational limit, we obtain Eq. (\ref{fa17}). At the minimum scattering length, we find that
\begin{eqnarray}
\label{fqw}
E_{tot}^*=-\frac{\nu^2}{3\sigma}\frac{G^2M^3m^2}{\hbar^2}.
\end{eqnarray}
The prefactor is $0.07074$. The total energy $E_{tot}$ is plotted as a function of the scattering length $a$ in Fig. \ref{ETOT-Mfixe} for a fixed value of the total mass $M$.

\subsection{The density profiles}
\label{sec_dpm}

\begin{figure}[!h]
\begin{center}
\includegraphics[clip,scale=0.3]{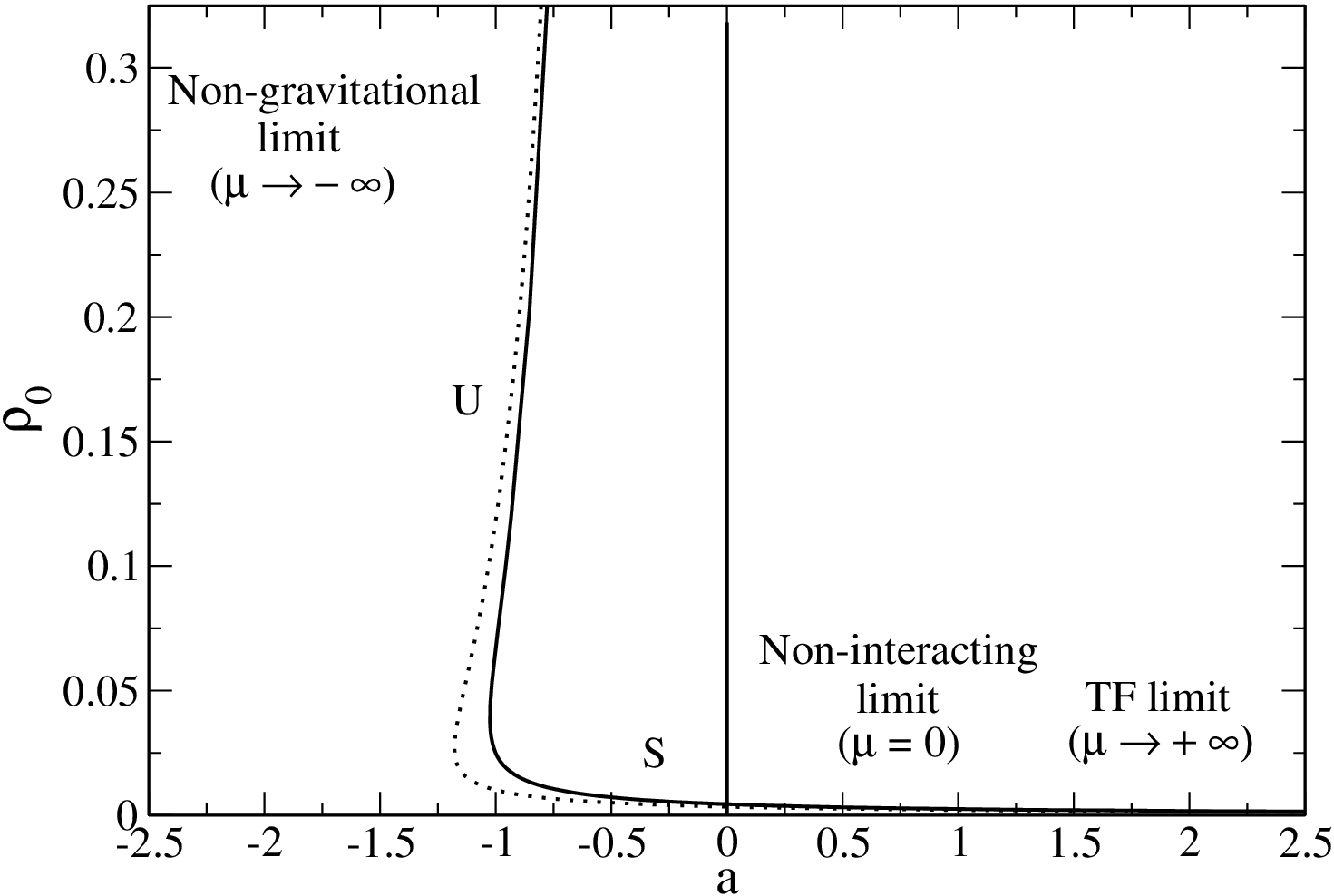}
\caption{$\rho_0$ as a function of $a$ for a fixed value of the mass $M$. The density is normalized by $\rho_Q$ and the scattering length by $a_Q$. In the non-interacting case $a=0$, we get $\rho_0\sim 4.400\, 10^{-3}$ and in the TF limit $a\rightarrow +\infty$, we get $\rho_0\sim 2.533\, 10^{-2}/a^{3/2}$. The central density at the point of minimum scattering length $a_{min}\simeq -1.025$ is $(\rho_0)_{*}\simeq 0.04$. We have also represented in dotted line the analytic expression obtained from the Gaussian ansatz. It is given by  ${\rho}_0=1/\pi^{3/2}{R}^3$ where $R$ is related to $a$ by the equation given in the caption of Fig. \ref{R99-part1}. This yields $\rho_0=(\nu^3/\pi^{3/2}\sigma^3)/(1\pm \sqrt{1+6\pi\zeta\nu a/\sigma^2})^3$. In the non-interacting case $a=0$ we get $\rho_0\sim (\nu^3/8\sigma^3\pi^{3/2})$ yielding $\rho_0^{Gauss}\sim 0.003378$, in the TF limit $a\rightarrow +\infty$ we get ${\rho}_0=({\nu}/{6\pi^2\zeta})^{3/2}a^{-3/2}$ yielding $\rho_0^{Gauss}\sim 0.03456/a^{3/2}$ and in the non-gravitational limit $a\rightarrow 0$ with $\rho_0\rightarrow +\infty$ we get $\rho_0\sim (\sigma/3\pi^{3/2}\zeta)^3/|a|^3$ yielding $\rho_0^{Gauss}\sim 0.3535/|a|^{3}$. Finally, the central density at the point of minimum scattering length is $(\rho_0)_*^{Gauss}=\nu^3/\pi^{3/2}\sigma^3=0.02703$. The approximate analytical curve has the same qualitative shape as the numerical curve and provides a good quantitative agreement for all values of the scattering length. The most severe deviation occurs close to the minimum scattering length value but it remains weak. }
\label{n0-part1-GAUSS}
\end{center}
\end{figure}

\begin{figure}[!h]
\begin{center}
\includegraphics[clip,scale=0.3]{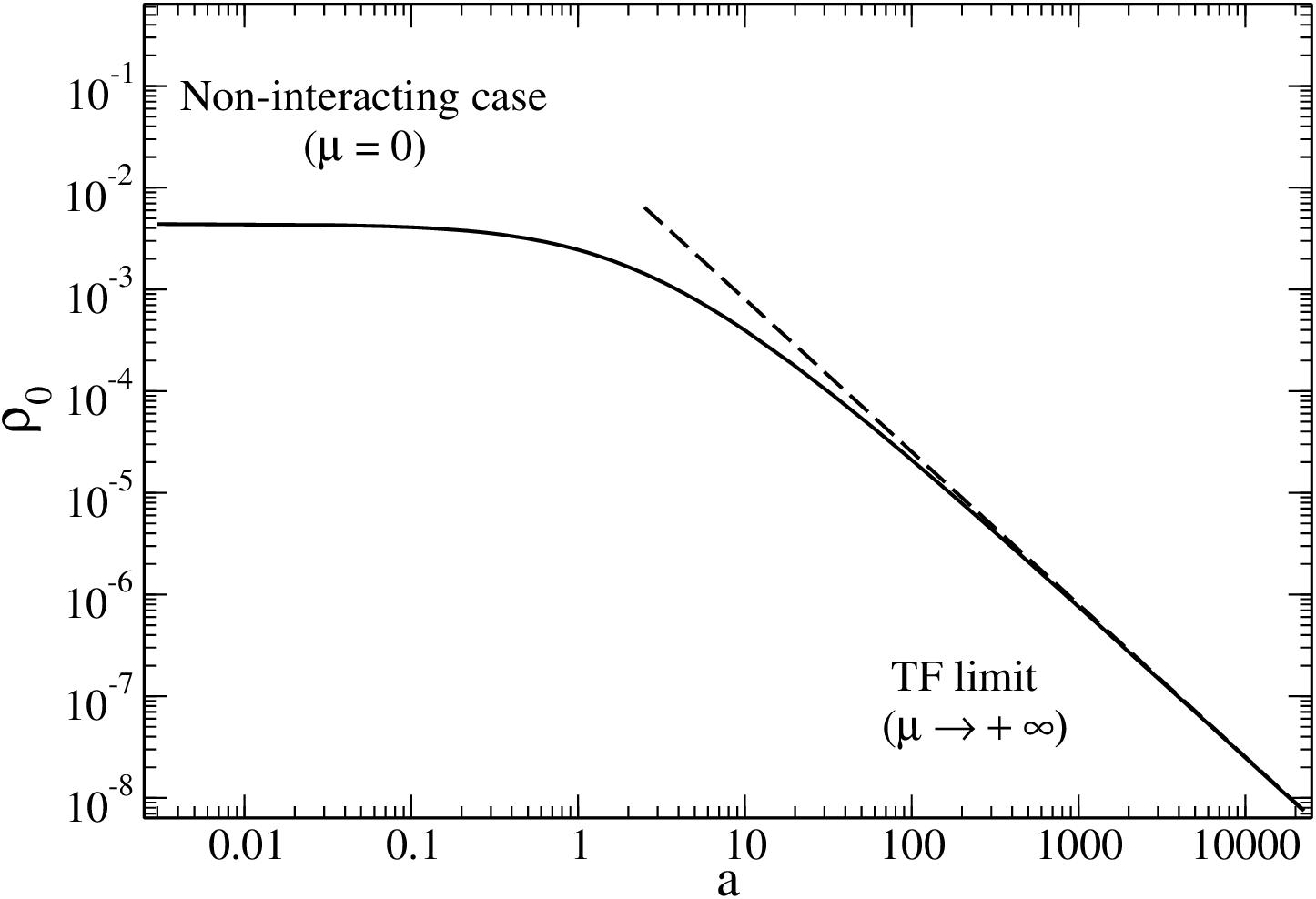}
\caption{$\rho_0$ as a function of $a$ (for $a>0$) in log-log plot for a fixed value of the mass $M$. The dashed line corresponds to $\rho_0\sim 2.533\, 10^{-2}/a^{3/2}$ characterizing  the TF limit.}
\label{n0-part2}
\end{center}
\end{figure}

The central density $\rho_0$ vs scattering length $a$ is plotted in Figs. \ref{n0-part1-GAUSS} and \ref{n0-part2} for a given value of the total mass $M$. In the noninteracting case $a=0$, it is given by Eq. (\ref{dp1}) and  in the TF limit $a\rightarrow +\infty$, it is given by Eq. (\ref{dp2}). The central density at the point of minimum scattering length is
\begin{equation}
\label{dpm1}
({\rho}_0)_*= 0.04\frac{G^3M^4m^6}{\hbar^6}.
\end{equation}

\begin{figure}[!h]
\begin{center}
\includegraphics[clip,scale=0.3]{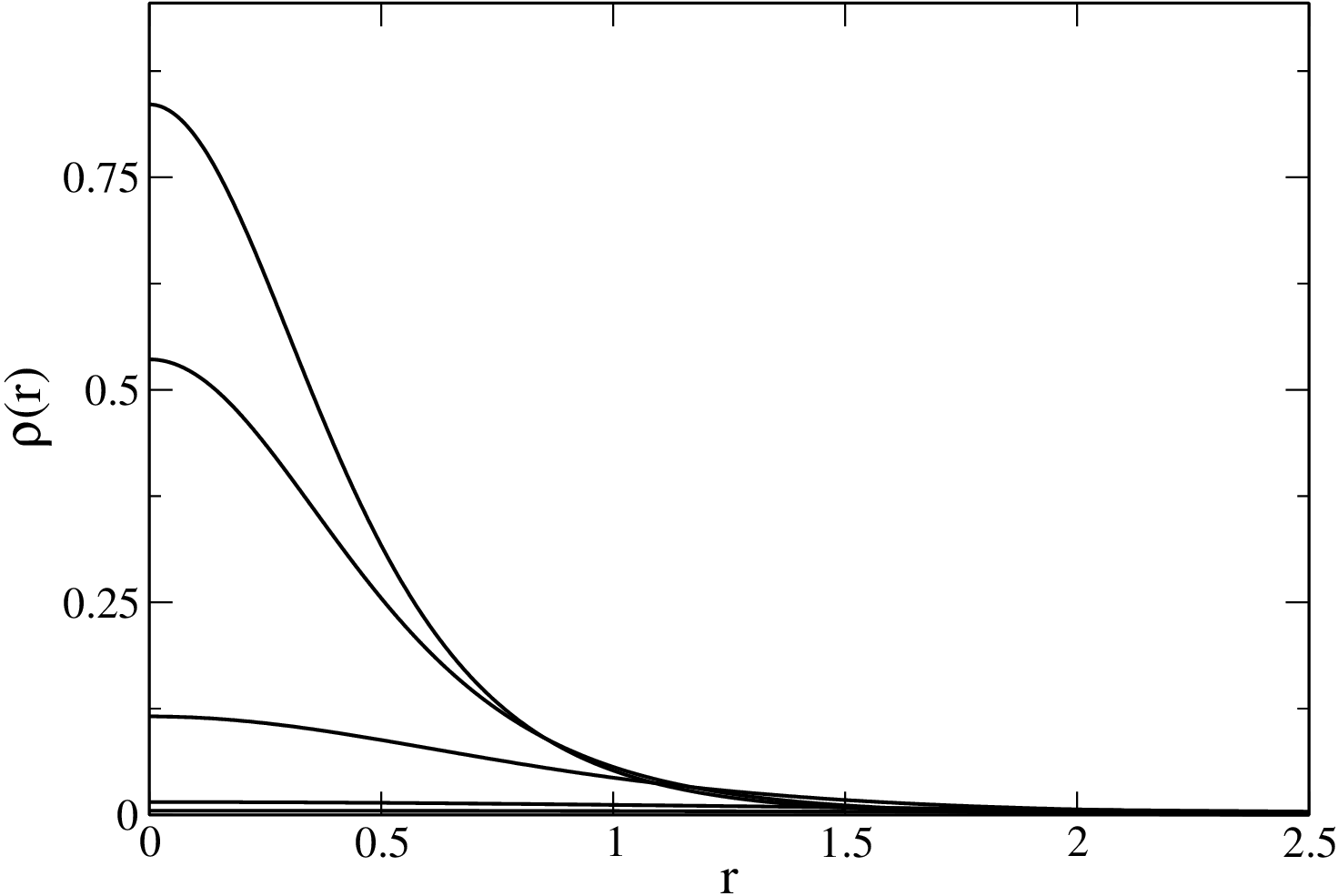}
\caption{Density profiles $\rho(r)$ along the series of equilibria for a fixed value of the total mass $M$. The density is normalized by $\rho_Q$ and the radial distance  by $R_Q$. The central density decreases as $\mu$ increases. The non-gravitational limit corresponds to $\mu\rightarrow -\infty$, the non-interacting case to $\mu=0$ and the TF limit to $\mu\rightarrow +\infty$. We have represented $\mu=-2.86, -2.55, -1.6, -0.55, -0.053$ (negative scattering lengths). The distribution becomes stable when $\mu>\mu_*\simeq -1.000$ corresponding to the turning point of scattering length (see Sec. \ref{sec_procedure}).}
\label{n-chi-neg}
\end{center}
\end{figure}

\begin{figure}[!h]
\begin{center}
\includegraphics[clip,scale=0.3]{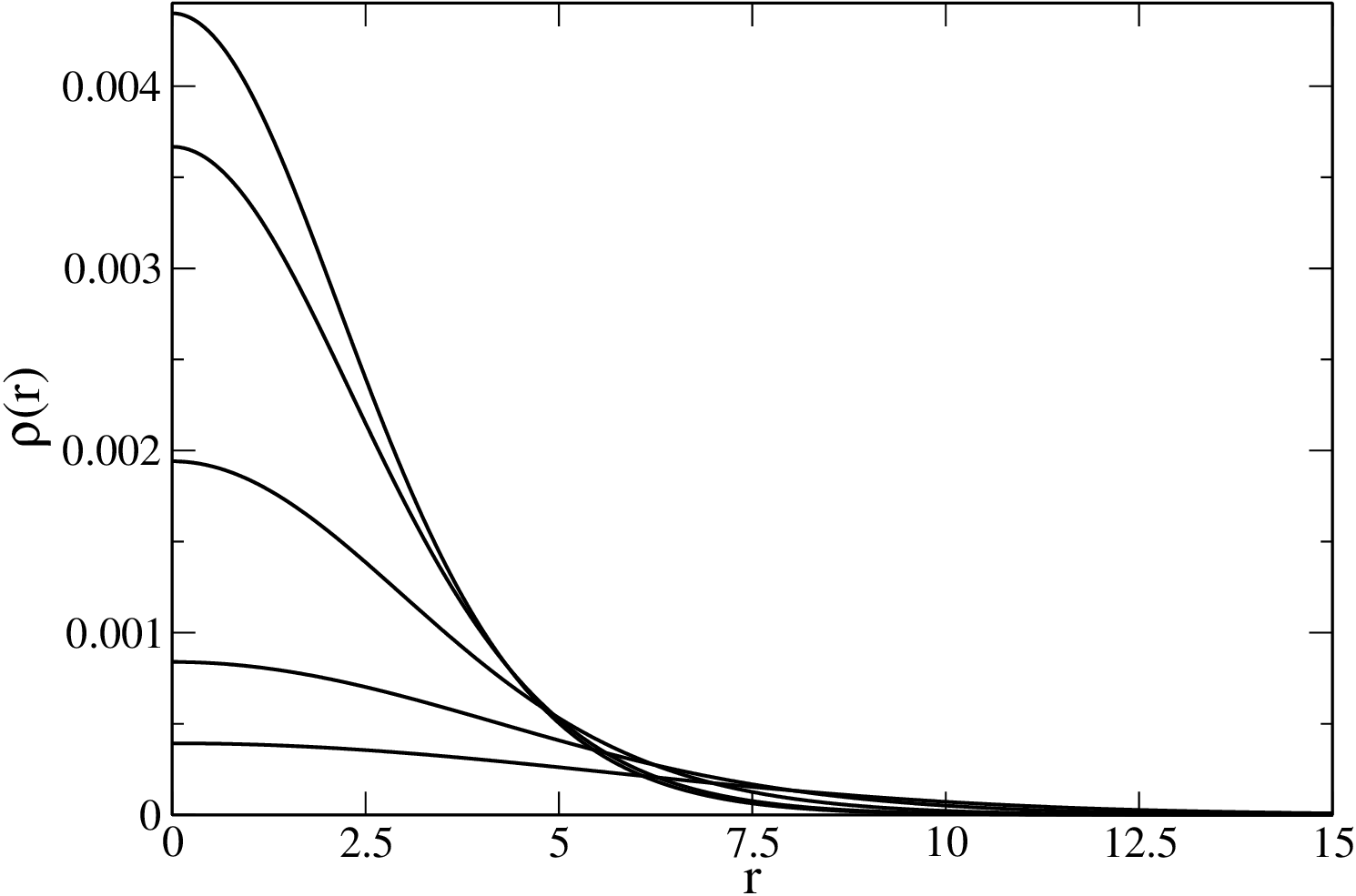}
\caption{Same as Fig. \ref{n-chi-neg} for  $\mu= 0, 0.078, 0.34, 0.69, 1$ (positive scattering lengths).}
\label{n-chi-pos-pt1}
\end{center}
\end{figure}

\begin{figure}[!h]
\begin{center}
\includegraphics[clip,scale=0.3]{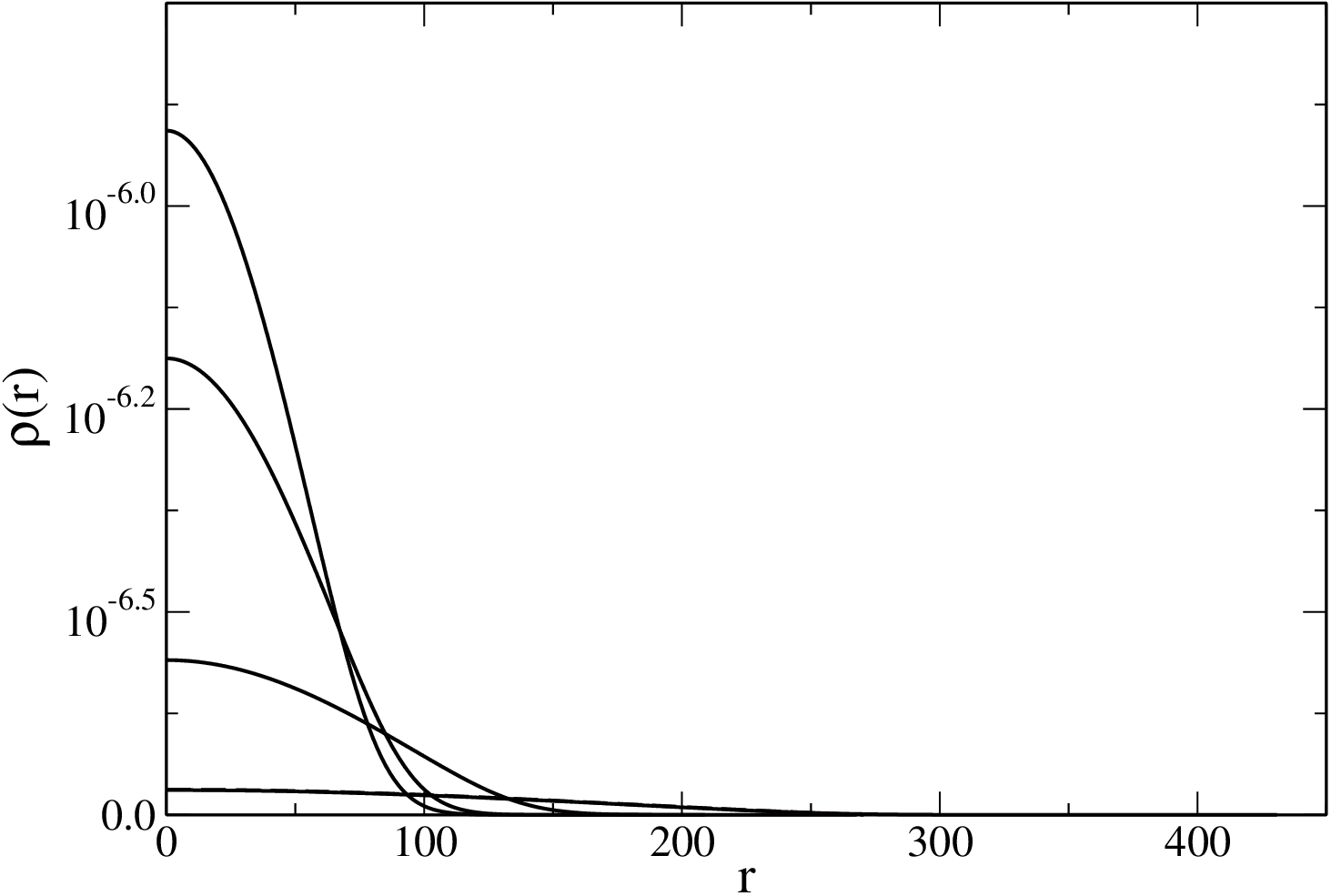}
\caption{Same as Fig. \ref{n-chi-neg} for  $\mu=4.1, 4.4, 5.3, 7.3$ (positive scattering lengths).}
\label{n-chi-pos-pt2}
\end{center}
\end{figure}

With the Gaussian ansatz, the central density is related to the mass and to the radius by Eq. (\ref{dp4}).
This equation gives the approximate central density versus radius relation for a fixed mass. Eliminating the radius between $a(R)$ and $\rho_0(R)$, we obtain $\rho_0$ as a function of $a$ in parametric form.  Alternatively, we can substitute the analytical expression  of $R(a)$ given by Eq. (\ref{fa5inv}) in Eq. (\ref{dp4}). In the non-interacting case $a=0$, we obtain Eq. (\ref{dp6}), in the Thomas-Fermi limit $a\rightarrow +\infty$ we obtain Eq. (\ref{dp7}) and in the non-gravitational limit we obtain Eq. (\ref{adp1}). Finally, the central density at the point of minimum scattering length is found to be
\begin{equation}
\label{dpm1g}
({\rho}_0)_*= \frac{\nu^3}{\pi^{3/2}\sigma^3}\frac{G^3M^4m^6}{\hbar^6}.
\end{equation}
This returns the scalings of Eqs. (\ref{dp1}), (\ref{dp2}) and (\ref{dpm1}) with the prefactors $0.003378$, $0.03456$ and $0.02703$ (the prefactor in the non-gravitational limit is $0.3535$). The analytical curve is compared with the numerical curve in Fig. \ref{n0-part1-GAUSS} and the agreement is fairly good. Some density profiles $\rho(r)$ are represented in Figs. \ref{n-chi-neg},  \ref{n-chi-pos-pt1}, \ref{n-chi-pos-pt2} for a fixed value of the total mass and different values of $a$.

\section{Conclusion}

In this paper, we have obtained the exact mass-radius relation of self-gravitating BECs with short-range interactions  by numerically solving the equation of hydrostatic equilibrium taking into account quantum effects. We have also compared our results with the approximate analytical relation obtained in Paper I from a Gaussian ansatz. We have found that the Gaussian ansatz always provides a good qualitative agreement with the exact solution, and that the agreement is quantitatively good in many cases. This gives us confidence to extend our approach to more general situations. This is interesting because analytical methods allow to explore a wider range of parameters than numerical methods and to obtain a more complete picture of the problem in parameter space. This will be the object of Paper III.

\appendix

\section{Numerical applications}
\label{sec_na}

In this Appendix, we try to relate our general results to real astrophysical systems and test the various approximations made in our study. To facilitate the comparison with astrophysical data, we express the radius $R$, the total mass $M$, the scattering length $a$ and the boson mass $m$ in terms of $M_{\odot}$, ${\rm kpc}$, ${\rm fm}$ and ${\rm eV}/c^2$.

\subsection{Fundamental quantities}
\label{sec_fq}

Let us first consider the case where the scattering length $a$ is fixed. The characteristic radius $R_a=(|a|\hbar^2/Gm^3)^{1/2}$ and the characteristic mass $M_a=\hbar/\sqrt{Gm|a|}$ can be written as
\begin{equation}
\label{app1}
\frac{R_a}{1{\rm kpc}}=5.558\, 10^{-3}\, \left (\frac{|a|}{1{\rm fm}}\right )^{1/2}\left (\frac{1 {\rm eV}/c^2}{m}\right )^{3/2},
\end{equation}
\begin{equation}
\label{app2}
\frac{M_a}{M_\odot}=1.537\, 10^{-34}\, \left (\frac{1{\rm fm}}{|a|}\right )^{1/2}\left (\frac{1 {\rm eV}/c^2}{m}\right )^{1/2}.
\end{equation}
For $a>0$, the characteristic radius $R_a$ corresponds to the minimum radius $R_{min}=\pi R_a$ of a Newtonian self-gravitating BEC, i.e. the radius of the BEC in the TF limit, and the characteristic mass $M_a$ corresponds to the typical mass separating the TF regime ($M\gg M_a$) from the non-interacting regime ($M\ll M_a$). For  $a<0$, the characteristic mass $M_a$ corresponds to the maximum mass $M_{max}=1.012 M_a$ and $R_a$ is the corresponding radius $R_{99}^*=5.5 R_a$.

Let us now consider the case where the total mass  $M$ is fixed. The characteristic radius $R_Q=\hbar^2/GMm^2$ and the characteristic scattering length $a_Q=\hbar^2/GM^2m$ can be written as
\begin{equation}
\label{app3}
\frac{R_Q}{1{\rm kpc}}=8.544\, 10^{-37}\, \frac{M_\odot}{M}\left (\frac{1 {\rm eV}/c^2}{m}\right )^{2},
\end{equation}
\begin{equation}
\label{app4}
\frac{|a_Q|}{1{\rm fm}}=2.363\, 10^{-68}\, \left (\frac{M_\odot}{M}\right )^{2}\frac{1 {\rm eV}/c^2}{m}.
\end{equation}
For $a\ge 0$, the characteristic radius $R_Q$ corresponds to the minimum radius $R_{99}^{min}=9.946R_Q$ of a Newtonian self-gravitating BEC, i.e. the radius of the BEC in the non-interacting limit, and the characteristic scattering length $a_Q$ corresponds to the typical scattering length separating the TF regime ($a\gg a_Q$) from the non-interacting regime ($a\ll a_Q$). For  $a<0$, the characteristic scattering length $a_Q$ corresponds to the minimum scattering length $a_{min}=-1.025 a_Q$ and $R_Q$ is the corresponding radius $R_{99}^*=5.6 R_Q$.

In order to make the correspondence between BECs with short-range interactions described by the Gross-Pitaevskii equation and scalar fields with a $\frac{1}{4}\lambda |\phi|^4$ interaction described by the Klein-Gordon equation, it is useful to introduce the dimensionless parameter \cite{paper1}:
\begin{equation}
\label{app5}
\frac{\lambda}{8\pi}\equiv \frac{a}{\lambda_c}=\frac{a m c}{\hbar},
\end{equation}
where $\lambda_c=\hbar/mc$ is the Compton wavelength of the bosons. It can be written as
\begin{equation}
\label{app6}
\frac{\lambda}{8\pi}=5.068\, 10^{-9}\, \frac{a}{1{\rm fm}}\frac{m}{1 {\rm eV}/c^2}.
\end{equation}
Using this expression, we can express the results in terms of $\lambda$ and $m$ instead of $a$ and $m$. In particular, we find that $R_a=(|\lambda|\hbar^3/8\pi Gm^4 c)^{1/2}$ and $M_a=(8\pi\hbar c/G|\lambda|)^{1/2}$ leading to
\begin{equation}
\label{app1b}
\frac{R_a}{1{\rm kpc}}=78.08\, \sqrt{\frac{|\lambda|}{8\pi}}\left (\frac{1 {\rm eV}/c^2}{m}\right )^{2},
\end{equation}
\begin{equation}
\label{app2b}
\frac{M_a}{M_\odot}=1.094\, 10^{-38}\, \sqrt{\frac{8\pi}{|\lambda|}}.
\end{equation}

Finally, we have seen in Paper I that the timescale of the BEC oscillations (for stable systems with $a>0$) or the collapse of the BEC (for unstable systems with $a<0$) corresponds to the dynamical time $t_D\sim (R^3/GM)^{1/2}$. It can be written
\begin{equation}
\label{td}
\frac{t_D}{1\, {\rm s}}=1.488\, 10^{19}\, \left (\frac{R}{1 {\rm kpc} }\right )^{3/2}\left (\frac{M_\odot}{M}\right )^{1/2}.
\end{equation}
On the other hand, the maximum circular velocity is $(v_c)_{max}\sim (GM/R)^{1/2}=R/t_D$. It can be written  
\begin{equation}
\label{vc}
\frac{(v_c)_{max}}{1\, {\rm km/s}}=2.074\, 10^{-3}\, \left (\frac{1 {\rm kpc}}{R}\right )^{1/2}\left (\frac{M}{M_\odot}\right )^{1/2}.
\end{equation}

\subsection{Orders of magnitude}
\label{sec_magnitude}

A specificity of our approach is to remain as general as possible (although restricting ourselves to the Newtonian approximation) so that we have expressed our results for an arbitrary value of the scattering length $a$ (positive, zero, or negative) and of the boson mass $m$. Therefore, our results can describe different objects, at different scales, such as (mini)-boson stars or galactic halos. In this subsection, and in the following subsections, we consider the case of galactic halos and make some numerical applications in order to check the validity of our approximations. To that purpose, we take a radius of order $R=10\, {\rm kpc}$ and a total mass of order $M=3\, 10^{11}M_\odot$ \cite{bohmer}. The corresponding dynamical time $t_D$ is of the order of $27$ Million years and the maximum circular velocity of order $360 \, {\rm km/s}$.

Let us first assume that the galactic halo can be modeled by a non-relativistic  gas of self-gravitating BECs with short-range interactions in the TF approximation. In that case, its radius $R=\pi R_a$ is determined by Eqs. (\ref{app1}) and (\ref{app1b}), and it is independent on the total mass. We have
\begin{equation}
\label{app1c}
\frac{R}{1{\rm kpc}}=0.01746\, \left (\frac{|a|}{1{\rm fm}}\right )^{1/2}\left (\frac{1 {\rm eV}/c^2}{m}\right )^{3/2},
\end{equation}
\begin{equation}
\label{app1d}
\frac{R}{1{\rm kpc}}=48.93\, \sqrt{\lambda}\left (\frac{1 {\rm eV}/c^2}{m}\right )^{2}.
\end{equation}
We see that the values of $a$ and $m$ given by B\"ohmer and Harko
\cite{bohmer}, namely $(a,m)=(1\, {\rm fm}, 14\, {\rm meV}/c^2)$ and
$(a,m)=(10^6\, {\rm fm}, 1.44\, {\rm eV}/c^2)$ give the correct order
of magnitude of the radius. For these values, $\lambda/8\pi=7.095\,
10^{-11}$ and $\lambda/8\pi=7.297\, 10^{-3}$ respectively. In fact, it
is important to realize that the radius $R$ of a self-coupled
Newtonian BEC directly determines the ratio $a/m^3$ or
$\lambda/m^4$. Using Eqs. (\ref{app1c}) and (\ref{app1d}) with $R=10\,
{\rm kpc}$, we obtain $m^3/a=3.049\, 10^{-6}({\rm eV/c^2})^3/{\rm fm}$
and $m^4/\lambda=23.94 \, ({\rm eV}/c^2)^4$ in agreement with the
estimate of Arbey {\it et al.} \cite{arbey} (they find a larger
numerical coefficient $\sim 50$ because they take a smaller halo
radius).

Let us now assume that the galactic halo can be modeled by a non-relativistic gas of self-gravitating BECs without short-range interaction. In that case, its  radius $R_{99}=9.946 R_Q$ is determined by Eq. (\ref{app3}) leading to
\begin{equation}
\label{app3b}
\frac{R_{99}}{1{\rm kpc}}=84.98\, 10^{-37}\, \frac{M_\odot}{M}\left (\frac{1 {\rm eV}/c^2}{m}\right )^{2}.
\end{equation}
We see that the mass of the bosons must be very small, of the order of  $m=10^{-24}\, {\rm eV}/c^2$, to reproduce the correct values of the radius and mass of the cluster. This boson mass corresponds to the estimate of Baldeschi {\it et al.} \cite{baldeschi} and others \cite{sin,hu,arbey1,matosall,silverman}.

Therefore, a self-interaction can increase the required value of the boson mass from $m=10^{-24}\, {\rm eV}/c^2$ to $m=1\, {\rm eV}/c^2$ which may be more realistic from a particle physics point of view.

\subsection{Validity of the TF approximation}
\label{sec_valtf}

For $a>0$, the validity of the TF approximation is determined by the dimensionless parameter \cite{paper1}:
\begin{equation}
\label{app7}
\chi=\frac{GM^2ma}{\hbar^2}.
\end{equation}
The TF approximation is valid for $\chi\gg 1$ while the non-interacting limit corresponds to $\chi\ll 1$. We can write $\chi=(M/M_a)^2=a/a_Q$. For a fixed value of $a$, the TF approximation is valid for $M\gg M_a$ and for a fixed value of $M$, the TF approximation is valid for $a\gg a_Q$. Equation (\ref{app7}) can be rewritten
\begin{equation}
\label{app8}
\chi=4.232\, 10^{67}\, \left (\frac{M}{M_\odot}\right )^2\frac{a}{1{\rm fm}}\frac{m}{1 {\rm eV}/c^2},
\end{equation}
or, equivalently,
\begin{equation}
\label{app8b}
\chi=8.361\, 10^{75}\, \left (\frac{M}{M_\odot}\right )^2\frac{\lambda}{8\pi}.
\end{equation}
Let us make a numerical application. If we want to model a galactic halo, then $M\simeq 3\, 10^{11}M_\odot$. Therefore, the TF approximation is valid if
\begin{equation}
\label{app9}
\frac{a}{1{\rm fm}}\frac{m}{1 {\rm eV}/c^2}\gg 2.626\, 10^{-91},
\end{equation}
or, equivalently,
\begin{equation}
\label{app10}
\frac{\lambda}{8\pi}\gg 1.331\, 10^{-99}.
\end{equation}
This relation clearly shows that {\it the limit $\lambda\rightarrow 0$ is different from the non-interacting case $\lambda=0$}. Indeed, the TF approximation is valid even for a (very) small value of $\lambda$ fulfilling the condition (\ref{app10}). By contrast, if $\lambda=0$ strictly, we are in the non-interacting case. In these two extreme limits, the physics of the problem is very different (see Appendix \ref{sec_magnitude} and Secs. II.E and II.F of Paper I). For the values of $a$ and $m$ given by B\"ohmer and Harko \cite{bohmer}, see Appendix \ref{sec_magnitude}, the condition (\ref{app10}) is fulfilled by more than $90$ orders of magnitude (!) so that the TF approximation is perfect.

\subsection{Validity of the Newtonian approximation}
\label{sec_valnewt}

The Newtonian approximation is valid if the radius $R$ of the configuration is much larger than the Schwarzschild radius $R_S=2GM/c^2$, or, equivalently, if
\begin{equation}
\label{app11}
M\ll \frac{R c^2}{G}.
\end{equation}
For fixed $a\neq 0$, this condition can be rewritten
\begin{equation}
\label{app12}
\frac{M}{M_a}\ll \kappa\frac{R}{R_a},\qquad {\rm with}\qquad \kappa=\frac{|a|c^2}{Gm}=\frac{|\lambda|\hbar c}{8\pi Gm^2}.
\end{equation}
The relativity parameter $\kappa$ can be written as
\begin{equation}
\label{app13}
\kappa=7.554\, 10^{47}\, \frac{|a|}{1{\rm fm}}\frac{1 {\rm eV}/c^2}{m},
\end{equation}
or, equivalently,
\begin{equation}
\label{app13b}
\kappa=1.491\, 10^{56}\, \frac{|\lambda|}{8\pi} \left (\frac{1 {\rm eV}/c^2}{m}\right )^2.
\end{equation}
We also note that $\kappa=(|\lambda|/8\pi)(M_P/m)^2$ where $M_P=(\hbar c/G)^{1/2}$ is the Planck mass. To determine the region in the $(M,R)$ plane of Figs. \ref{M-R-chi-pos-part1} and \ref{M-R-chi-neg-part1} where the Newtonian approximation is valid, it suffices to draw a straight line with slope $\kappa$. The Newtonian approximation is valid below this line. In general, $\kappa$ is large (see Eqs. (\ref{app13}) and \ref{app13b})) so that the Newtonian approximation is a good approximation as soon as $a\neq 0$, even if it is (very) small. Let us try to be more specific by considering particular limits.

In the TF limit, the radius of the classical self-gravitating BEC is of the order $R_a$. Using Eq. (\ref{app11}), we find that the Newtonian approximation is valid if
\begin{equation}
\label{app14}
M\ll \frac{\hbar c^2 a^{1/2}}{(Gm)^{3/2}}\sim \sqrt{\frac{\lambda}{8\pi}}\frac{1}{m^2}\left (\frac{\hbar c}{G}\right )^{3/2}.
\end{equation}
Using Eqs. (\ref{app1}), (\ref{app6}) and (\ref{app12}), this condition can be rewritten
\begin{equation}
\label{app15}
\frac{M}{M_{\odot}}\ll 1.161\, 10^{14}\, \left (\frac{a}{1{\rm fm}}\right )^{1/2}\left (\frac{1 {\rm eV}/c^2}{m}\right )^{3/2},
\end{equation}
or, equivalently,
\begin{equation}
\label{app16}
\frac{M}{M_{\odot}}\ll 1.631\, 10^{18}\, \sqrt{\frac{\lambda}{8\pi}}\left (\frac{1 {\rm eV}/c^2}{m}\right )^{2}.
\end{equation}
If we want to model a galactic halo, recalling the numerical values of $M$, $a$ and $m$ given in Appendix \ref{sec_magnitude}, we see that the term in the left hand side is of order $10^{11}$ while the term in the right hand side is of order $10^{17}$. Therefore, the condition (\ref{app15})-(\ref{app16}) is fulfilled by $6$ orders of magnitude so that the Newtonian approximation is very good. We note that the mass appearing in the right hand side of Eq. (\ref{app14}) represents the maximum mass of a relativistic self-gravitating BEC with short-range interactions in the TF limit (up to a numerical factor that can be obtained by solving the general relativistic equation of hydrostatic equilibrium \cite{chavharko}). As shown in Appendix B3 of Paper I, it corresponds to the maximum mass $M_{max}=0.062 \sqrt{\lambda}M_P^3/m^2$ found by Colpi {\it et al.} \cite{colpi}. The corresponding radius is still given by $R_a$ which can be written $R=0.3836 \sqrt{\lambda}(M_P/m)\lambda_c$ or $R=6.187 GM_{max}/c^2$ \cite{chavharko}. We have
\begin{equation}
\label{app15b}
\frac{M_{max}}{M_{\odot}}= 1.141\, \left (\frac{a}{1{\rm fm}}\right )^{1/2}\left (\frac{1 {\rm GeV}/c^2}{m}\right )^{3/2},
\end{equation}
\begin{equation}
\label{app16b}
\frac{M_{max}}{M_{\odot}}= 0.1011\, \sqrt{\lambda}\left (\frac{1 {\rm GeV}/c^2}{m}\right )^{2},
\end{equation}
\begin{equation}
\label{rads}
{R}= 9.272\, \frac{M_{max}}{M_\odot} \, {\rm km}.
\end{equation}
For $m\sim 1{\rm GeV}/c^2$, $a\sim 1{\rm fm}$ and $\lambda\sim 1$, the maximum  mass $M_{max}\sim M_{\odot}$ is of the order of the solar mass, and the corresponding radius $R$ is of the order of the kilometer, like in the case of white dwarf and neutron stars. This could describe boson stars with relevant mass. We emphasize that the mass $M$ (or the radius $R$) of the  self-coupled relativistic boson stars directly  determine the ratio $a/m^3$ or $\lambda/m^4$. Taking  $M=1\, M_\odot$,  we obtain 
$m^3/a=1.302 \, ({\rm GeV/c^2})^3/{\rm fm}$ and  $m^4/\lambda=0.01023 \, ({\rm GeV}/c^2)^4$.

In the non-interacting case ($a=0$), the radius of the classical self-gravitating BEC is of the order of $R_Q$. Using Eq. (\ref{app11}), we find that the Newtonian approximation is valid if
\begin{equation}
\label{app17}
M\ll \frac{\hbar c}{Gm}, \qquad {\rm or}\qquad R\gg \frac{\hbar}{mc}.
\end{equation}
Using Eqs.  (\ref{app3}) and (\ref{app12}), this condition can be rewritten
\begin{equation}
\label{app18}
\frac{M}{M_\odot}\ll 1.336\,  10^{-10}\, \frac{1 {\rm eV}/c^2}{m},
\end{equation}
or
\begin{equation}
\label{app19}
\frac{R}{1{\rm kpc}}\gg 6.395\, 10^{-27}\, \frac{1 {\rm eV}/c^2}{m}.
\end{equation}
If we want to model a galactic halo, recalling the numerical values of $M$ and $m$ given in Appendix \ref{sec_magnitude}, we see that the term in the left hand side of Eq. (\ref{app18}) (resp. Eq. (\ref{app19})) is of order $10^{11}$ (resp. $10$) while the term in the right hand side is of order $10^{14}$ (resp. $10^{-2}$). Therefore, the conditions (\ref{app18}) and (\ref{app19}) are fulfilled by $3$ orders of magnitude so that the Newtonian approximation is very good. We note that the mass and the radius appearing in the right hand side of Eq. (\ref{app17}) represent the maximum mass and the minimum radius of a relativistic self-gravitating BEC without short-range interaction. As shown in Appendix B.2 of Paper I, it corresponds to the Kaup mass $M_{Kaup}=0.633 M_P^2/m$ and Kaup radius $R_{95}=6.03 \lambda_c$ or $R=9.526 GM_{max}/c^2$. We have
\begin{equation}
\label{app18b}
\frac{M_{Kaup}}{M_\odot}= 0.8457\,  10^{-10}\, \frac{1 {\rm eV}/c^2}{m},
\end{equation}
\begin{equation}
\label{app19b}
R_{95}= 14.07 \frac{M_{Kaup}}{M\odot}\, {\rm km}.
\end{equation}
For $m\sim 1{\rm GeV}/c^2$, the Kaup mass $M_{Kaup}\sim 10^{-19}M_{\odot}$ is very small and irrelevant. The Kaup mass becomes of the order of the solar mass if the bosons have a mass $m\sim 10^{-10}{\rm eV}/c^2$ (leading to a radius of the order of the km). For example, axionic boson stars could account for the mass of MACHOs (between $0.3$ and $0.8$ $M_{\odot}$) if the axions have such a small mass \cite{mielkeschunck}.

Finally, in the case of attractive short-range interactions ($a<0$), the Newtonian approximation is valid if $\kappa\gg (M_{max}/M_a)/(R_{99}^*/R_a)=0.184$, a condition that is easily fulfilled in practice (see Eqs. (\ref{app13}) and (\ref{app13b})). As shown in Appendix B.4 of Paper I, we have $M_{max}=5.073 M_P/\sqrt{|\lambda|}$ and $R_{99}^*=1.1  \sqrt{|\lambda|}(M_P/m)\lambda_c$.

\end{document}